# New Constraints on the Evolution of the $M_{\rm H\,I}$–$M_\star$ Scaling Relation Combining CHILES and MIGHTEE-H I Data

Alessandro Bianchetti[1,2], Francesco Sinigaglia[3,4], Giulia Rodighiero[1,2], Ed Elson[5], Mattia Vaccari[6,7,8], D.J. Pisano[9], Nicholas Luber[10], Isabella Prandoni[8], Kelley Hess[11,12], Maarten Baes[13], Elizabeth A.K. Adams[12,14], Filippo M. Maccagni[15], Alvio Renzini[2], Laura Bisigello[1,2], Min Yun[16], Emmanuel Momjian[17], Hansung B. Gim[16,18], Hengxing Pan[19], Thomas A. Oosterloo[12,17], Richard Dodson[20], Danielle Lucero[21], Bradley S. Frank[6,9,22,23], Olivier Ilbert[24], Luke J.M. Davies[20], Ali A. Khostovan[25], and Mara Salvato[26]

[1] Department of Physics and Astronomy, Università degli Studi di Padova, Vicolo dell'Osservatorio 3, I-35122, Padova, Italy
[2] INAF—Osservatorio Astronomico di Padova, Vicolo dell'Osservatorio 5, I-35122, Padova, Italy
[3] Département d'Astronomie, Université de Genève, Chemin Pegasi 51, 1290 Versoix, Switzerland
[4] Institut für Astrophysik, Universität Zürich, Winterthurerstrasse 190, CH-8057 Zürich, Switzerland
[5] Department of Physics and Astronomy, University of the Western Cape, Robert Sobukwe Road, 7535 Bellville, Cape Town, South Africa
[6] Inter-university Institute for Data Intensive Astronomy, Department of Astronomy, University of Cape Town, 7701 Rondebosch, Cape Town, South Africa
[7] Inter-university Institute for Data Intensive Astronomy, Department of Physics and Astronomy, University of the Western Cape, 7535 Bellville, Cape Town, South Africa
[8] INAF—Istituto di Radioastronomia, via Gobetti 101, 40129 Bologna, Italy
[9] Department of Astronomy, University of Cape Town, Private Bag X3, Rondebosch 7701, South Africa
[10] Department of Astronomy, Columbia University, Mail Code 5247, 538 West 120th Street, New York, NY 10027, USA
[11] Department of Space, Earth and Environment, Chalmers University of Technology, Onsala Space Observatory, SE-43992 Onsala, Sweden
[12] ASTRON, the Netherlands Institute for Radio Astronomy, Oude Hoogeveesedijk 4, 7991 PD Dwingeloo, The Netherlands
[13] Sterrenkundig Observatorium, Universiteit Gent, Krijgslaan 281 S9, 9000 Gent, Belgium
[14] Kapteyn Astronomical Institute, University of Groningen, P.O. Box 800, 9700 AV Groningen, The Netherlands
[15] INAF—Osservatorio Astronomico di Cagliari, via della Scienza 5, 09047, Selargius (CA), Italy
[16] Department of Astronomy, University of Massachusetts, Amherst, MA 01003, USA
[17] National Radio Astronomy Observatory, P.O. Box O, Socorro, NM 87801, USA
[18] Department of Physics, Montana State University, P.O. Box 173840, Bozeman, MT 59717, USA
[19] Oxford Astrophysics, Denys Wilkinson Building, University of Oxford, Keble Road, Oxford, OX1 3RH, UK
[20] International Centre for Radio Astronomy Research, The University of Western Australia, 35 Stirling Highway, Crawley, WA, Australia
[21] Department of Physics, Virginia Polytechnic Institute and State University, 50 West Campus Drive, Blacksburg, VA 24061, USA
[22] STFC UK Astronomy Technology Centre, Royal Observatory, Edinburgh, Blackford Hill, Edinburgh, EH9 3HJ, UK
[23] South African Radio Astronomy Observatory (SARAO), 2 Fir Street, Observatory, 7925, South Africa
[24] Aix Marseille Universit'e, CNRS, CNES, LAM, Marseille, France
[25] Laboratory for Multiwavelength Astrophysics, School of Physics and Astronomy, Rochester Institute of Technology, 84 Lomb Memorial Drive, Rochester, NY 14623, USA
[26] Max Planck Institute for Extraterrestrial Physics, Giessembachstrasse 1, D-857498, Garching, Germany

Received 2024 December 10; revised 2025 January 24; accepted 2025 January 31; published 2025 March 20

## Abstract

The improved sensitivity of interferometric facilities to the 21 cm line of atomic hydrogen (H I) enables studies of its properties in galaxies beyond the local Universe. In this work, we perform a 21 cm line spectral stacking analysis combining the MeerKAT International GigaHertz Tiered Extragalactic Exploration and COSMOS H I Large Extra-galactic Survey surveys in the COSMOS field to derive a robust H I–stellar mass relation at $z \approx 0.36$. In particular, by stacking thousands of star-forming galaxies subdivided into stellar mass bins, we optimize the signal-to-noise ratio of targets and derive mean H I masses in the different stellar mass intervals for the investigated galaxy population. We combine spectra from the two surveys, estimate H I masses, and derive the scaling relation $\log_{10} M_{\rm HI} = (0.32 \pm 0.04)\log_{10} M_\star + (6.65 \pm 0.36)$. Our findings indicate that galaxies at $z \approx 0.36$ are H I richer than those at $z \approx 0$ but H I poorer than those at $z \approx 1$, with a slope consistent across redshift, suggesting that stellar mass does not significantly affect H I exchange mechanisms. We also observe a slower growth rate H I relative to the molecular gas, supporting the idea that the accretion of cold gas is slower than the rate of consumption of molecular gas to form stars. This study contributes to understanding the role of atomic gas in galaxy evolution and sets the stage for future development of the field in the upcoming Square Kilometre Array era.

*Unified Astronomy Thesaurus concepts:* Galaxy formation (595); Galaxy evolution (594); Emission line galaxies (459); Radio astronomy (1338); Extragalactic astronomy (506); Extragalactic radio sources (508)

## 1. Introduction

Currently, only ~10% of cosmic baryons have been converted into stars (P. Madau & M. Dickinson 2014). The baryon cycle encompasses the set of mechanisms of gas exchange that connect the various components of a galaxy with its surrounding environment. These include how gas collapses to form stars, how feedback triggers and regulates new star formation, and how neutral and molecular gas reservoirs are consumed and replenished in a delicate balance. Understanding the interaction among these components—ranging from stars to various phases of the interstellar medium—is essential to comprehend the formation and evolution of galaxies.

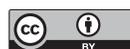

In this context, we aim to explore the role of neutral atomic hydrogen (H I). H I acts as a precursor to molecular hydrogen (H$_2$), which is the primary fuel for star formation. A common approach to investigate the role of H I in galaxy evolution involves parameterizing scaling relations that link H I to other key properties of galaxies, such as stellar mass ($M_\star$), the star formation rate (SFR), the specific star formation rate (sSFR), galaxy size, and color.

Recent studies of the nearby Universe suggest that regions with intense star formation are associated with larger amounts of H I (S. Huang et al. 2012; Z. Zhou et al. 2018), mirroring the well-established correlation between cold gas mass and SFR (R. C. J. Kennicutt 1998; F. Bigiel et al. 2008; S. Roychowdhury et al. 2009, 2015). Contextually, the quenching of star formation in red galaxies is often attributed to the low fraction of H I, among other contributing factors. In fact, star formation may cease when H I reservoirs cannot replenish the molecular phase (A. Saintonge & B. Catinella 2022). However, the relation between atomic gas and star formation is more complex than it may seem. For instance, approximately 30% of early-type galaxies in the *Atlas3D* sample (M. Cappellari et al. 2011; P. Serra et al. 2012), which includes objects with different morphologies across the full mass spectrum, show evidence of H I content. This indicates that other factors beyond the mere quantity of cold gas significantly influence the regulation of star formation.

It is also widely accepted that H I is removed from galaxies through multiple environment-dependent processes, whereas inflowing streams originate from extragalactic reservoirs such as the circumgalactic medium, the intergalactic medium, or filamentary structures (R. Sancisi et al. 2008; C. L. Carilli & F. Walter 2013; F. Walter et al. 2020). Such environmental effects are typically observed in galaxies at scales larger than the optical radius, where H I halos dominate and serve as useful tracers for tidal interaction or ram pressure stripping (H. Dénes et al. 2016; L. Cortese et al. 2021, and references therein).

Numerous studies have investigated the interplay between molecular and atomic gas in galaxies. Notably, L. Morselli et al. (2021) suggest that the $M_{H_2}/M_{H\,I}$ ratio within the optical radius of star-forming sources slightly decreases with increasing redshift ($\propto (1 + z)^{-0.34}$), which is contrary to the common expectation that galaxies become progressively more dominated by molecular hydrogen at high redshifts (e.g., L. J. Tacconi et al. 2018, 2020; T.-M. Wang et al. 2022). Therefore, it is crucial to explore the H I content in galaxies beyond the local Universe.

Scaling relations are tools to improve our understanding of the connections between the various processes and mass components in a galaxy. In this work, we investigate the scaling relation between H I mass and stellar mass in star-forming galaxies. This relation has been extensively studied at $z \sim 0$ (S. Huang et al. 2012; N. Maddox et al. 2015; H. Guo et al. 2021, hereafter G21; J. Rhee et al. 2023), thanks to the availability of rich data sets from large-scale H I galaxy surveys, such as the H I Parkes All-Sky Survey (D. G. Barnes et al. 2001), the Arecibo Legacy Fast ALFA Survey (ALFALFA; R. Giovanelli et al. 2005), the Deep Investigation of Neutral Gas Origins (DINGO; M. Meyer 2009), and the GALEX Arecibo SDSS Survey (GASS; B. Catinella et al. 2010). H I properties of galaxies have been further studied up to $z \sim 0.2$ and beyond through various untargeted surveys, including the Blind Ultra-Deep H I Environmental Survey (M. Verheijen et al. 2007; Y. L. Jaffé et al. 2016), the Arecibo Ultra-Deep Survey (L. Hoppmann et al. 2015), the COSMOS H I Large Extra-galactic Survey (CHILES; X. Fernández et al. 2013, 2016; K. M. Hess et al. 2019; J. Blue Bird et al. 2020; R. Dodson et al. 2022) and the MeerKAT International GigaHertz Tiered Extragalactic Exploration (MIGHTEE; M. Jarvis et al. 2016). The comprehensive findings from these collaborative efforts reveal a complex picture in which H I properties correlate with several galaxy parameters, such as stellar mass, luminosity, morphology, size, SFR, and environment (F. Bigiel et al. 2008, 2010; B. Catinella et al. 2013; A. Boselli et al. 2014; S. Janowiecki et al. 2017; D. Kleiner et al. 2017; A. Saintonge & B. Catinella 2022; H. Pan et al. 2023; F. Sinigaglia et al. 2024). Nevertheless, it is still not conclusively established which among these correlations are the most fundamental and which most significantly drive the complex phenomenology of the baryon cycle (L. Lin et al. 2019). Thanks to modern interferometric facilities such as the Westerbork Synthesis Radio Telescope, the Karl G. Jansky Very Large Array (VLA) at the National Radio Astronomy Observatory, the Giant Metrewave Radio Telescope (GMRT), MeerKAT, and other Square Kilometre Array (SKA) precursors, the angular resolution of 21 cm line observations has significantly improved, enabling the study of resolved objects beyond the local Universe. Moreover, the high sensitivity and large collecting areas in single-dish radio facilities helped radio astronomers to extract H I information at higher redshift (e.g., the Five-hundred-meter Aperture Spherical Radio Telescope; R. Nan et al. 2011). However, direct detection of H I in single galaxies is still limited to local galaxies or very massive galaxies at redshift $z < 0.5$, due to the intrinsic faintness of the 21 cm line.

To overcome this issue, a powerful alternative is the cost-effective observational technique known as spectral line stacking (e.g., M. A. Zwaan 2000; J. Delhaize et al. 2013; Q. Chen et al. 2021), and it can be used when direct H I detections cannot be achieved. This technique enables statistical measurement of the average H I mass ($M_{H\,I}$) of a given galaxy sample. Stacking has been proven to be a powerful tool in studying various aspects of galaxy evolution, including the investigation of H I abundance in galaxy clusters (e.g., P. Lah et al. 2009; J. Healy et al. 2021), the $M_{H\,I}$ content of AGN host galaxies (K. Geréb et al. 2015; F. M. Maccagni et al. 2017), cold gas stripping in satellite galaxies (T. Brown et al. 2017), and the redshift evolution of the H I cosmic density parameter (e.g., J. Delhaize et al. 2013; J. Rhee et al. 2013; A. Chowdhury et al. 2020, and references therein). Specifically, J. Rhee et al. (2013, 2016, 2018) reported tentative H I statistical detections (i.e., at a significance $<3\sigma$) at $z \approx 0.2$, $z \approx 0.37$, and $z \approx 0.32$, respectively, using stacking methods. A. Bera et al. (2019) report a stacking detection at $z \sim 0.34$, while F. Sinigaglia et al. (2022, hereafter S22) and A. Bera et al. (2023a) present the $M_{H\,I}-M_\star$ and $M_{H\,I}-$SFR scaling relations at $z \sim 0.4$ again based on stacking technique. A. Chowdhury et al. (2020, 2021) generated stacked H I detections at $z \sim 1$ and $z \sim 1.3$, respectively. A. Chowdhury et al. (2022, hereafter C22) derived the highest-redshift reference for the $M_{H\,I}-M_\star$ scaling relation ($z \sim 1$), approaching the so-called "cosmic noon"—when the SFR density reaches its peak (P. Madau & M. Dickinson 2014).

In this paper, we employ a spectral stacking pipeline (S22; F. Sinigaglia et al. 2024) to statistically recover the H I signal below the nominal sensitivity of two state-of-the-art blind H I surveys. In particular, we aim to revisit the scaling relations obtained in the COSMOS field from the MIGHTEE Large Survey Program at $z \approx 0.37$ (S22), as well as to measure the same relation from CHILES at the same redshift and within a smaller field completely encircled in COSMOS, and finally to combine them as independent data sets to obtain a robust global scaling relation.

This paper is structured as follows. In Section 2, we present the details of the MIGHTEE survey and CHILES, characterize the data we use for stacking herein, and describe the sample selection. In Section 3, we describe in detail the standard stacking procedure and explain how we combine MIGHTEE and CHILES. Section 4 presents the combined stacks together along with the best-fit linear relation and contextualizes them within the broader literature at various redshifts. In Section 5, we discuss and interpret our findings. Finally, we summarize the main points of the discussion and present our conclusions in Section 6.

Throughout the work, we assume a spatially flat ($\Omega_k = 0$) $\Lambda$CDM cosmology, employing cosmological parameters derived from the latest Planck collaboration results (Planck Collaboration et al. 2020), i.e., $H_0 = 67.4$ km s$^{-1}$ Mpc$^{-1}$, $\Omega_m = 0.315$, and $\Omega_\Lambda = 0.685$, and an initial mass function as in G. Chabrier (2003).

## 2. The MIGHTEE Survey and CHILES

The MeerKAT radio interferometer (J. Jonas & MeerKAT Team 2016) is located in South Africa and consists of 64 offset Gregorian dishes equipped with receivers in the UHF band (580 MHz $< \nu <$ 1015 MHz), $L$ band (900 MHz $< \nu <$ 1670 MHz), and $S$ band (1750 MHz $< \nu <$ 3500 MHz) and serves as a precursor to the SKA, whose full operations are expected around 2030. MIGHTEE is an $L$-band continuum, polarization, and spectral line large survey conducted with MeerKAT, utilizing spectral and full Stokes mode observations. It covers four deep extragalactic fields (COSMOS, XMM-LSS, ECDFS, and ELAIS-S1), chosen because of their extensive multi-wavelength coverage from previous and ongoing observations. For this paper, we use the MIGHTEE-H I Early Science spectral line data from MIGHTEE (N. Maddox et al. 2021). The observations were carried out between mid-2018 and mid-2019 around the COSMOS field (N. Scoville et al. 2007), covering a total area of $\sim 5$ deg$^2$ at $z = 0$. Data were processed using the processMeerKAT calibration pipeline (J. D. Collier et al. 2021), which implements standard calibration routines and strategies, including flagging, bandpass, and complex gain calibration. Continuum subtraction was performed in the visibility domain using CASA routines *uvsub* and *uvcontsub*. The resulting data cubes underwent median filtering to reduce artifacts associated with direction-dependent errors (N. Maddox et al. 2021).

This study uses MIGHTEE-H I Early Science data cubes covering the COSMOS field with a single pointing within the redshift range of $0.22 < z < 0.49$, with an effective exposure time of approximately 23 hr. The beam is approximately $17\rlap{.}''2 \times 13\rlap{.}''9$ at $z \sim 0.36$ ($\approx 90$ kpc $\times$ 73 kpc). The median H I noise rms of the cubes increases as the frequency decreases, ranging from 85 $\mu$Jy beam$^{-1}$ at $\nu \approx 1050$ MHz to 135 $\mu$Jy beam$^{-1}$ at $\nu \approx 950$ MHz at a spectral resolution (channel width) of 209 kHz (roughly 80 km s$^{-1}$ at $z = 0.36$ using the optical definition of velocity). The spectral bands corresponding to redshift ranges $0.09 < z < 0.22$ and $z > 0.49$ are excluded from the analysis due to strong radio frequency interference (RFI) features (N. Maddox et al. 2021).

CHILES is a deep-field H I survey, carried out with the VLA and imaging H I over a contiguous redshift range of $0 < z < 0.49$ for the first time (X. Fernández et al. 2016). The pointing was centered on R.A. (J2000) $10^h 01^m 24^s$ and decl. (J2000) $2^d 21^m 00^s$. It samples a subregion of the COSMOS field; therefore, it provides an independent measurement of the same patch of sky as the MIGHTEE survey. The 25 m diameter antennas of the VLA provide a field of view of around $0\rlap{.}°5$ at 1.4 GHz, imaged into 1201 $\times$ 1201 pixels channel$^{-1}$. The VLA-B configuration was used, with baselines up to 11 km and a typical beam size of approximately $7\rlap{.}''2 \times 6\rlap{.}''4$ at $z \approx 0.36$ ($\approx 38 \times 33$kpc$^2$). The data cube comprises 125 kHz wide spectral channels (approximately 50 km s$^{-1}$ at $z \approx 0.36$ using the optical definition of velocity), covering a wide range of frequencies (950–1420 MHz). The rms ranges between 30 and 50 $\mu$Jy beam$^{-1}$ in this range. Due to the rotating VLA configurations, the CHILES observations were split into five observing epochs spaced approximately 15 months apart. In this paper, we use successfully processed data from all five of the observing epochs, amounting to approximately 800 observation hr (approximately 600 on-source hr), for which details on the calibration can be found in D. J. Pisano et al. (2025, in preparation), and imaged using the technique in N. Luber et al. (2025, in preparation). Fifteen spectral windows, each 32 MHz wide, were utilized, resulting in a total of 480 MHz per observing session, each having continuous exposure. To ensure continuous frequency coverage without gaps, daily observations were dithered, with each observation epoch having its specific frequency dithering settings.

The MIGHTEE and CHILES footprints are shown in Figure 1, where we used a slice from the MIGHTEE-H I (N. Maddox et al. 2021) Early Science cubes, at $z \approx 0.36$. We mark the area imaged by CHILES with a brown dashed box. The footprint of CHILES is completely contained within that of MIGHTEE. Blue and red dashed circles provide a visual indication of the primary beam sizes for MIGHTEE and CHILES, respectively.

### 2.1. Sample Selection

We cross-match the information obtained from photometric observations with a robust spectroscopic catalog, as highly accurate redshifts are needed for stacking. We only select star-forming galaxies in the COSMOS field, within the redshift range simultaneously covered by both data cubes ($0.22 < z < 0.49$): within this interval, the MIGHTEE-H I Early Science cubes do not suffer significantly from RFI.

#### 2.1.1. Spectroscopic Sample

The spectroscopic catalog we use consists of an updated and higher-quality version of the spectroscopic redshift catalog used in S22, constructed by merging three different catalogs covering the COSMOS field: the COSMOS spec-$z$ compilation (A. Khostovan et al. 2025, in preparation), the DEVILS survey catalog (L. J. M. Davies et al. 2018), and the DESI survey catalog (DESI Collaboration et al. 2024). We introduce a quality cut on the first of the aforementioned compilations by selecting only those spectroscopic redshifts that have a confidence level ⩾80%. As for DESI, only redshifts from the primary catalog with no warnings (*ZCAT_PRIMARY* and *ZWARN* = 0) are used. For DEVILS, we select galaxies marked by *SpecFlag* > 2.

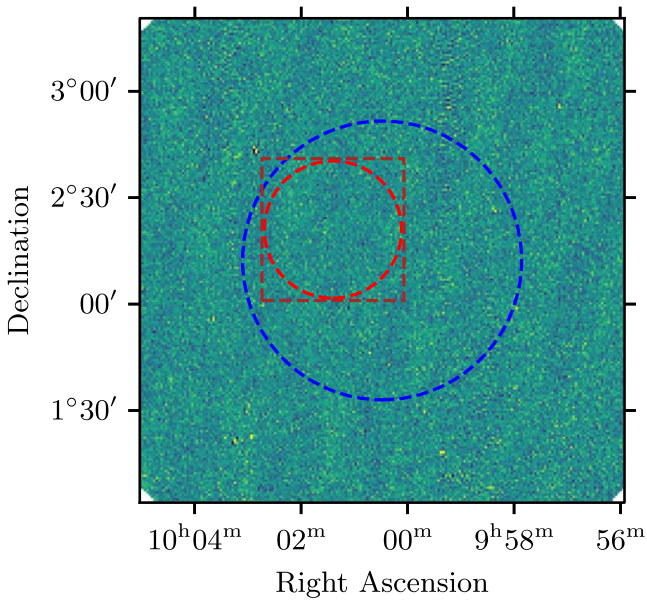

**Figure 1.** Single-channel map from the MIGHTEE-H I Early Science data cube extracted at frequency $f \approx 1036$ MHz ($z \approx 0.36$). The brown dashed rectangle marks the area imaged in the CHILES data cube. The blue dashed circle marks the region where the primary beam (PB) model of MeerKat is equal to 0.5. The red dashed circle marks the same limit for the VLA and is positioned at the center of CHILES.

### 2.1.2. Photometric Sample and Physical Properties

Photometric information is retrieved from the latest publicly available COSMOS photometric sample from the COSMOS2020-CLASSIC data release (J. R. Weaver et al. 2022).[27] We perform a positional cross-matching (matching radius = 1″) between the photometrically derived parameters from *LePhare* and our merged master spec-z catalog. The photometric catalog includes derived galaxy properties (specifically, stellar mass, SFR, and photometric redshift) obtained through spectral energy density (SED) fitting. We then identify star-forming galaxies using a color–color selection in the $(NUV - r)/(r - J)$ rest-frame plane, using the same criterion as in C. Laigle et al. (2016). Quiescent galaxies are defined as those that meet the conditions $NUV - r > 3(r - J) + 1$ and $NUV - r > 3.1$, while the remaining galaxies are classified as star-forming. This criterion is less sensitive to reddening than using the $U - V$ color and also reduces the mixing of quiescent and dusty star-forming galaxies (we refer the reader to O. Ilbert et al. 2013 for the details). Still, *UVJ*-selected quiescent samples have 21%–30% contamination from galaxies with significant levels of ongoing star formation (e.g., C. Schreiber et al. 2018; B. Forrest et al. 2020). As SED fitting is based on photometric redshift estimates, we identify outliers in the determination of photometric redshift following the criterion $|z_{phot} - z_{spec}| > 0.15(1 + z_{spec})$ (T. Dahlen et al. 2013) and exclude them from the sample. This operation classifies about 10% of the galaxies as outliers and yields a sample of 8314 star-forming, spectroscopically selected galaxies in the $0.22 < z < 0.49$ range, of which 2486 lie in the CHILES footprint.

---

[27] Release 1, v2.2, 2023 March.

Figure 2 shows the normalized distributions of galaxy number counts for spectroscopic redshift (top left), stellar mass (top right), SFR (bottom left), and sSFR (bottom right) for the final galaxy sample, subdivided into the MIGHTEE (blue) and the CHILES (red) data sets. We use normalized counts to make visual comparison more straightforward. The median value and profiles of all the distributions from the two surveys are in excellent agreement, supporting the argument that the two samples are characterized by very similar physical properties. One of the main factors possibly causing minor discrepancies in the comparison between the two histograms is cosmic variance, since the two sets of galaxies are drawn from regions in the sky with a different cosmic volume. We investigate whether cosmic variance has repercussions on the final results in Appendix A.2.

Overall, J. R. Weaver et al. (2022, 2023) have shown that the COSMOS photometric sample is complete in stellar mass down to $\log_{10} M_\star / M_\odot \approx 8$ at $z < 0.5$. Figure 2 (top right panel) highlights a drop in the number of sources for $\log_{10} M_\star / M_\odot \lesssim 9$ due to spectroscopic selection. This means that the spectroscopic sample shows evidence of completeness only for $\log_{10} M_\star / M_\odot \gtrsim 9$; therefore, the cross-matched sample will inherit this property. The observed incompleteness at lower stellar masses is not surprising, since the master spectroscopic catalog was assembled from a combination of several different surveys. However, the impact of incompleteness should be mitigated by the fact that we are splitting the sample into stellar mass bins. Specifically, we identify the four following bins:

1. $8.0 < \log M_\star / M_\odot < 9.5$,
2. $9.5 < \log M_\star / M_\odot < 9.8$,
3. $9.8 < \log M_\star / M_\odot < 10.5$,
4. $\log M_\star / M_\odot > 10.5$.

These bins are not defined based on the criterion of containing approximately the same number of galaxies; rather, they are selected heuristically to contain enough galaxies to obtain a detection of the H I signal with a robust signal-to-noise ratio (S/N) value in each bin.

Moreover, we verify that the distribution of the spectroscopically selected sample in the $M_\star$–SFR plane qualitatively exhibits a symmetric distribution around the main sequence (or MS, the main sequence of star-forming galaxies; G. Rodighiero et al. 2014; J. S. Speagle et al. 2014). In particular, Figure 3 presents our sample together with the MS parameterization at $z = 0.36$ as modeled by P. Popesso et al. (2023). This is fundamental to demonstrate that the color criterion we employed successfully selects star-forming galaxies and that we are statistically sampling the MS. However, there is a slight deviation at the higher masses, where a significant fraction of massive galaxies lies below the MS. We assume that this is a consequence of contamination in the color selection by massive galaxies on their way to being quenched. To quantify this effect, we assume a reference dispersion around the MS of ∼0.6 dex as a criterion to assign membership (G. Rodighiero et al. 2011) and find that about 25% of galaxies in this bin lie below the lower limit given by the scatter. We thus exclude such galaxies from the sample to ensure that we are still properly sampling the MS even in the high-mass bin. The average values of stellar mass and SFR after the cut are shown in Figure 3 (magenta stars).

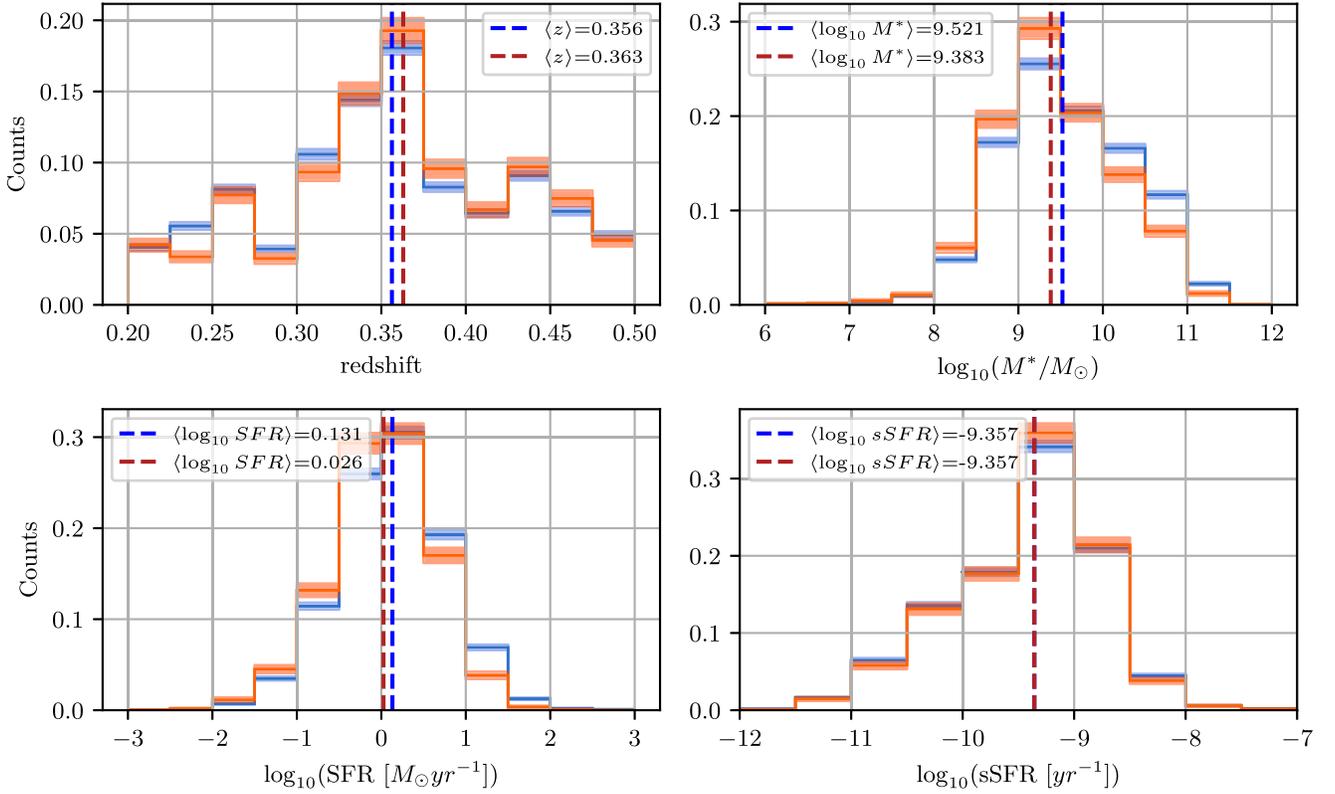

**Figure 2.** Physical properties of the sample at $\langle z \rangle = 0.36$: normalized histograms of redshift (top left), stellar mass (top right), SFR (bottom left), and sSFR (bottom right). We display the distributions related to MIGHTEE and CHILES in blue and red, respectively. We also assign an uncertainty to each bin, given by the Poisson shot noise. The dashed vertical line indicates the median value for the two surveys. In all cases, there is good consistency between the distribution of the full MIGHTEE sample (on the whole COSMOS field) and the CHILES subsample.

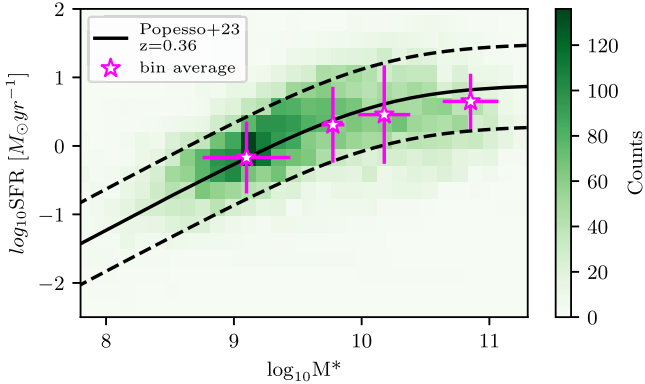

**Figure 3.** 2D histogram of stellar mass vs. SFR for the sample at $0.22 < z < 0.49$. A model for the MS from P. Popesso et al. (2023) at $z = 0.36$ is also superimposed (black solid line). The black dashed lines mark a 0.6 dex scatter estimate (G. Rodighiero et al. 2011). Magenta stars represent the average $\log M_\star$ and $\log SFR$ for each of the stellar mass bins that will be used for stacking.

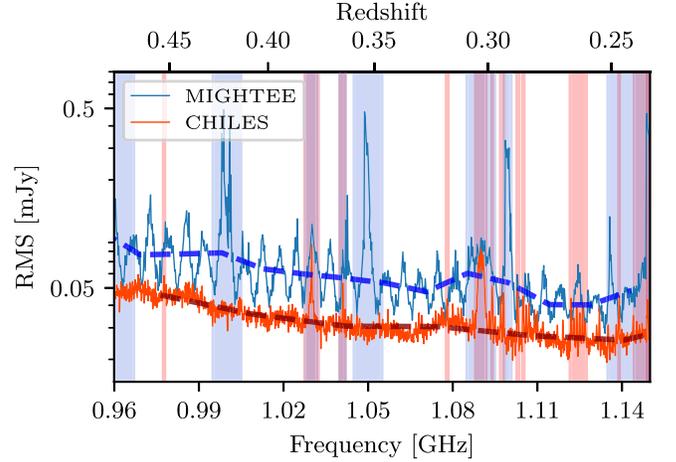

**Figure 4.** rms as a function of frequency for the MIGHTEE (blue) and CHILES (red) data cubes analyzed in this work. In each channel of a data cube, we extract 100 randomly placed, beam-sized apertures; compute the integrated flux densities within the apertures; and compute the rms of these 100 values. The blue and dark red dashed lines represent the median trend of the noise for MIGHTEE and CHILES, respectively. The two surveys have different spectral resolutions: to account for the mismatch, we downscaled the noise trend of CHILES by a factor $F \sim 1.3$ in this plot (see text). The shaded vertical bands pinpoint the noisy windows identified in the RFI flagging procedure. Galaxies falling in these frequency ranges are excluded from the stacking procedure.

## 3. Stacking Methodology

For this analysis, we use a stacking routine (S22; F. Sinigaglia et al. 2024) based on a standard procedure for spectral line stacking (see, for example, J. Healy et al. 2019; Q. Chen et al. 2021). As described in F. Sinigaglia et al. (2024), we include an RFI masking technique by excluding frequency bands that are significantly noisy, assuming that they are affected by RFI. To do that, we analyze the variation of rms as a function of frequency, which we show in Figure 4. The rms value is obtained by computing the variance among the integrated flux densities within 100 randomly placed beams within the sky map in each channel. For a fair comparison, the rms is given in units of Jy instead of Jy beam$^{-1}$, thereby removing the dependence on the beam size. The noise profile for the MIGHTEE cube displays some noticeable periodic peaks: the highest four spikelike features correspond to the boundaries of the artificial spectral windows imposed by the data reduction pipeline. The same figure also shows that the CHILES rms level shows fewer features and stays just slightly below the MIGHTEE rms level in terms of average flux sensitivity. Consecutive channels presenting high rms values are interpreted as RFI-affected frequency windows. These are marked by shaded areas in Figure 4. Galaxies falling in these frequency ranges are excluded from the sample, leading to a ∼15% reduction in the sample size.

If a galaxy falls inside an RFI-free region, we extract a H I cubelet that surrounds its location in the data cube. The center of each cubelet is defined using the optical coordinates and spectroscopic redshift from our merged spectroscopic catalog. The extracted cubelet spans 3× the FWHM of the beam major axis in the sky domain (R.A.–decl.) while extending approximately ±2000 km s$^{-1}$ along the spectral axis, here given in velocity units. The angular aperture used to extract the cubelets (3 times the beam FWHM) represents a conservative choice to encompass the entirety of the emitted H I flux from galaxies—extending beyond their optical size—independently of whether they are resolved or not. Three times the CHILES beam (the smallest of the two) corresponds to a physical scale of ≈100 kpc at $z = 0.36$, which is comparable with the typical size of larger H I disks (see J. Wang et al. 2016; S. H. A. Rajohnson et al. 2022, for references on the $M_{\rm H\,I}$–size relation at $z = 0$).

Next, we integrate each cubelet across the angular coordinates to obtain a spectrum. Each spectrum, initially observed at frequency $\nu_{\rm obs}$, is then shifted to its rest-frame frequency $\nu_{\rm rf}$ and converted to velocity units using $v = cz$. Given that the frequency bin width is constant throughout the data cubes ($\Delta f = 209$ kHz for MIGHTEE and $\Delta f = 125$ kHz for CHILES, corresponding to ∼80 km s$^{-1}$ and ∼50 km s$^{-1}$ at $z \sim 0.36$, respectively), the velocity bin widths vary with redshift. To maintain uniform spectral binning, spectra are resampled on a reference spectral template with a fixed velocity bin width of $\Delta v = 100$ km s$^{-1}$, ensuring flux conservation.

The conversion of spectra from flux density units to $M_{\rm H\,I}$ in units of $M_\odot$ per km s$^{-1}$ follows

$$M_{\rm HI}(v) = (2.356 \times 10^5) D_L^2\, S(v)(1+z)^{-1}\, M_\odot\, {\rm km}^{-1}\, {\rm s} \quad (1)$$

from M. S. Roberts (1962), where $D_L$ represents the luminosity distance of the galaxy in Mpc units, $S(\nu)$ is the 21 cm spectral flux density in Jy, and $(1+z)^{-1}$ corrects for flux reduction due to the expansion of the Universe.

At this stage, we apply a further quality cut. We first compute the mean value of the flux from the channels of the spectrum, excluding the channels falling within the integration region of the stacked spectrum (whose width is discussed at the end of this subsection). Then, we compare the mean to the normal distribution of the means of all stacked spectra and exclude a spectrum from the sample if its mean flux exceeds the 3σ limit, as explained in Appendix A.1. This cut ensures that we eliminate any spectra that deviate from the Gaussian distribution of flux densities.

The selected spectra are coadded, resulting in a stacked spectrum described by the equation

$$\langle M_{\rm HI}(v)\rangle = \frac{\sum_{i=0}^{n} M_{{\rm HI},i}(v) \times w_i \times f_i}{\sum_{i=0}^{n} w_i \times f_i^2}. \quad (2)$$

Here, $n$ denotes the number of coadded spectra, while $f_i$ and $w_i$ represent the average transmission of the primary beam and the weight assigned to each source, respectively. To model $w_i$, in an attempt to down-weight noisy spectra, we use an inverse power of a proxy for the noise level, taken to be the rms over the channels outside the integration region. In practice, we adopt $w_i = 1/\sigma_i^\gamma$, where we use $\gamma = 1$ in our standard approach, as in S. Fabello et al. (2011). In addition, the value $\gamma = 1$ is also the choice that maximizes S/N in the stacking procedure applied to the MIGHTEE data used (F. Sinigaglia et al. 2024). Equation (2) implements the primary beam correction based on the procedure described in K. Geréb et al. (2013). We also assign a 1σ uncertainty to the H I mass estimate obtained by computing the rms of the channels in the stacked spectrum, i.e., those that lie outside the integration range. For this paper, we choose an integration range of ±350 km s$^{-1}$, since this interval is found to maximize the S/N and should encompass the typical width of the stacked signal (Section 4).

Finally, we quantify the integrated S/N of the final stacked spectrum as the ratio between the mass estimate and the σ value defined above. This is expressed as

$$S/N = \sum_{i}^{N_{\rm ch}} \langle S_i \rangle / (\sigma \sqrt{N_{\rm ch}}), \quad (3)$$

where $\sum_{i}^{N_{\rm ch}} \langle S_i \rangle$ is the integrated flux density of the stacked spectrum and $N_{\rm ch}$ is the number of channels over which the integration is performed. This assesses the statistical significance of our measurements.

### 3.1. Combined Stacking

The two surveys were conceived for different scientific purposes: MIGHTEE is wider and shallower (≈23 effective hr for the Early Science Data), covering the full COSMOS field (∼2 deg$^2$), while CHILES (1000 hr) covers a narrower area and is deeper. These characteristics make the two surveys complementary and allow us to probe a deep sample with robust statistics, which is fundamental for our stacking purposes.

Figure 4 shows a comparison of the trend of noise rms as a function of frequency of the MIGHTEE (blue) and CHILES (red) cubes. The noise baselines for the two surveys are a factor of ∼2.3 apart on average. We note that the cubes have different spectral resolutions ($\Delta f_C = 125$ kHz for CHILES, $\Delta f_M = 209$ kHz for MIGHTEE-HI Early Science). The radiometer equation states that rms $\propto \Delta f^{-1/2}$, where $\Delta f$ is the frequency window. As we will work with the homogenizing bin width in spectra later in the analysis, we introduced a downscaling factor $F = \sqrt{\Delta f_M / \Delta f_C}$ and applied it to the CHILES curve, to account for the difference in spectral resolution.

In preparation for the combined stacking, we also probe the behavior of noise with respect to the number of stacked spectra separately for the two different cubes. To this end, we extract 5000 cubelets at random R.A., decl., and $z$ coordinates distributed uniformly within the data cube and then stack them. At each step of the stacking procedure, we compute the rms of the stacked spectrum. For both data cubes, we expect that the rms of the noise of a stacked spectrum scales $\propto \sqrt{N}$, where $N$ is the number of stacked spectra. Therefore, rms/$N \propto N^{-1/2}$. Figure 5 displays the result of a stack in flux density (mJy), where the noise agrees well with the expected trend for both MIGHTEE and CHILES and appears to be very stable. This provides an important step to validate the stacking procedure. The uncertainties displayed are obtained by computing the variance of $N$ realizations of cubelets extracted at random positions.

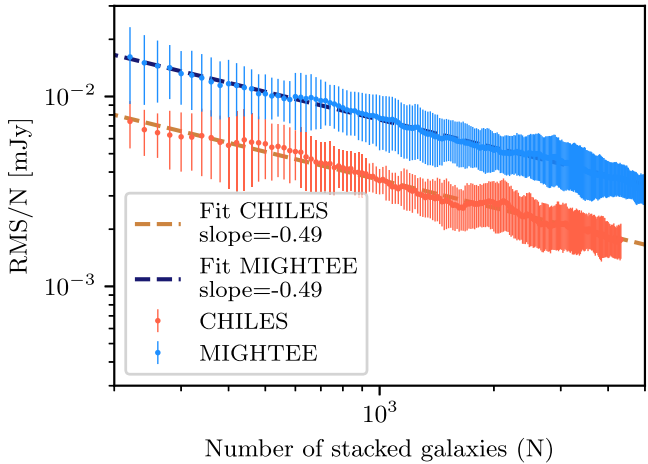

**Figure 5.** Noise rms normalized to the number of stacked spectra $N$ as a function of $N$. The curves are obtained at each $N$ by measuring the rms of the flux densities of all channels of the stacked spectrum (see the text for further details). The uncertainties are estimated through a Monte Carlo sampling using the standard deviation of 100 realizations of the sample. Finally, we fit the rms/$N$ trend with a power law (rms/$N \propto N^\theta$) and find that both for CHILES and for MIGHTEE, the power-law index $\theta$ is consistent with $-1/2$, as expected.

To support the validity of the combined stacking approach, we test the consistency between the stacked spectra on the two different surveys, starting from a sample defined by the same physical parameters. This is discussed in Appendix A.2. In general, we find good agreement between the stacking results of the two data sets. In particular, the H I mass estimates are found to be consistent in all three probed stellar mass bins.

Operationally, we will use the full MIGHTEE footprint to extract spectra: if a galaxy falls inside the region overlapping with CHILES, we will have two spectra, which we will treat as two separate, independent instances. Since spectra from two different surveys are extracted at different native spectral resolutions, we need to resample them to a common reference system. This is automatically performed during the conversion from frequency to velocity units when all the spectra (even if extracted from the same cube) are rebinned following the same spectral template, which makes it straightforward to then stack them. As mentioned in Section 3, we resample the spectra to a common bin width of $100 \, \text{km s}^{-1}$.

### 3.2. Confusion Correction

We outline here the strategy to address the issue of source confusion, i.e., flux contamination due to unknown or known nearby companions of the stacked galaxies. S22 studied source confusion using detailed MeerKAT-like simulated data cubes configured with the same technical specifics as MIGHTEE-H I Early Science observations and injecting realistic H I sources constructed from a mock galaxy catalog (D. Obreschkow & M. Meyer 2014), following the methodology outlined in E. C. Elson et al. (2016).

We refer to S22 for the details and assume the same contamination level in the present study. The estimated confusion contribution to the total flux is $\sim 10\%$ at $\langle z \rangle = 0.36$, which we apply as a correction to our measurements. We also refer the reader to E. C. Elson et al. (2019) for a detailed simulation-based study of source confusion as a function of redshift and based on realistic observational scenarios.

## 4. Results

In this section, we present the results of our stacking experiments.

### 4.1. $M_{H\,I}$–$M_\star$ Relation from Combined Stacking

We present here the results from the combination of the two surveys. By gathering larger statistics from these two independent data sets, we can distribute the stellar mass range into a finer grid than the case of a single survey, which in turn increases the robustness of the stellar-to-neutral hydrogen mass scaling relation of MS galaxies at $z \approx 0.4$. The latter point is crucial to better constrain the scaling relation and understand more in depth its evolution with cosmic time.

For each of the aforementioned stellar mass bins ($8.0 < \log M_\star/M_\odot < 9.5$, $9.5 < \log M_\star/M_\odot < 9.8$, $9.8 < \log M_\star/M_\odot < 10.5$, and $\log M_\star/M_\odot > 10.5$), we run the stacking pipeline, including the RFI masking procedure and adopting a weighting scheme as in S. Fabello et al. (2011; $w_i = 1/\sigma_i$). We show the stacks obtained in the four bins in Figure 6. All four stacks display a relatively narrow peak (width $\lesssim 700 \, \text{km s}^{-1}$). The continuum features some slight fluctuations, which may consist in the residual of the continuum subtraction process in the MIGHTEE-H I Early Science data and/or of non-Gaussian artifacts. However, we correct for this effect by fitting a second-order polynomial function to the continuum (excluding the spectral range where the line is located) and subtracting it from the baseline. The width of the 21 cm line in the stacked spectra ($\Delta v \approx \pm 500 \, \text{km s}^{-1}$) is greater than the typical width measured from the single spectra ($\Delta v \approx \pm 250$–$300 \, \text{km s}^{-1}$). This fact is mainly due to the uncertainty in the spectroscopic redshifts ($\Delta v \lesssim 100 \, \text{km s}^{-1}$; e.g., S. J. Lilly et al. 2007), which smears the double-horn profile and broadens the line (e.g., N. Maddox et al. 2013; E. C. Elson et al. 2019).

The extracted neutral hydrogen masses and other bin parameters are listed in Table 1, where we report S/N > 5 in all bins. Table 1 shows that different bins have a similar redshift distribution, since their median values coincide within 5%. Thus, we exclude any redshift-related selection effect. The last column ($M_{H\,I,\,\text{corr}}$) lists the final H I masses adjusted for the aforementioned 10% confusion factor (Section 3.2).

In Figure 7, we report the H I masses extracted in each bin as a function of the average stellar mass value in the corresponding bin (blue squares).

We fit the four resulting data points with a power law. The mean values for the fitting parameters and the associated uncertainties (68% confidence level) are obtained using a parametric bootstrap of the data, generating $10^4$ samples, and fitting the model with a least-squares minimization for each sample. Thus, we obtain the following best-fitting relation, displayed in Figure 7 (blue solid line):

$$\log_{10} M_{\text{HI}} = (0.32 \pm 0.04)\log_{10} M_\star + (6.65 \pm 0.36). \quad (4)$$

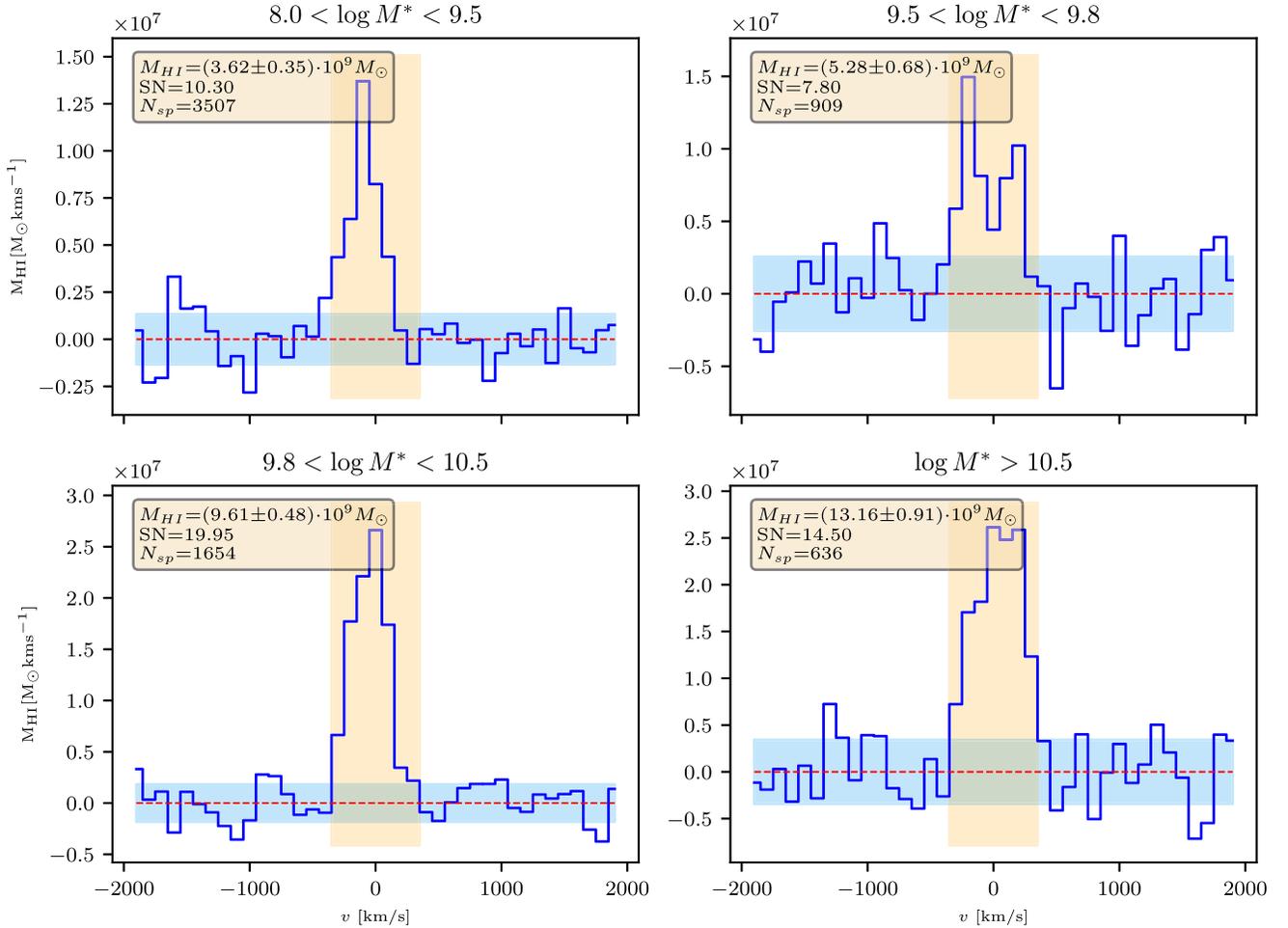

**Figure 6.** $M_{\rm H\,I}$ stacks in the four studied stellar mass bins. The used bin width is $\Delta v = 100$ km s$^{-1}$. The yellow shaded area represents the integration range in the rest-frame velocity domain, spanning from $-350$ km s$^{-1}$ to $+350$ km s$^{-1}$. The red dotted line marks the $M_{\rm H\,I} = 0$ line. For each panel, we report the extracted hydrogen mass and its associated uncertainty, estimated as the rms of the bins in the region of the stacked spectrum lying outside the integration range (cyan shaded region). Finally, we report the S/N and the number of spectra selected in that stellar mass bin.

**Table 1**
Summary of the Stacking Results for MIGHTEE and CHILES Combined at $\langle z \rangle = 0.36$

| $\overline{M_\star}$ [$\times 10^9 M_\odot$] | $N_{\rm sp}$ | $\langle z \rangle$ | $M_{\rm H\,I}$ [$\times 10^9 M_\odot$] | S/N | $M_{\rm H\,I,\,corr}$ [$\times 10^9 M_\odot$] |
|---|---|---|---|---|---|
| 1.1 | 3507 | 0.350 | $3.62 \pm 0.35$ | 10.3 | $3.26 \pm 0.35$ |
| 4.4 | 909 | 0.370 | $5.28 \pm 0.68$ | 7.8 | $4.75 \pm 0.68$ |
| 13.1 | 1546 | 0.362 | $9.61 \pm 0.48$ | 20.0 | $8.64 \pm 0.48$ |
| 56.6 | 636 | 0.371 | $13.15 \pm 0.91$ | 14.5 | $12.84 \pm 0.91$ |

**Note.** For each stellar mass bin, we report its mean value (first column), the number of coadded spectra (second column), the median redshift of the galaxies in that bin (third column), the integrated neutral hydrogen mass $M_{\rm H\,I}$ (fourth column), the S/N of the stack (fifth column), and the hydrogen mass corrected for confusion (sixth column). Note that $M_{\rm H\,I}$ was extracted in the velocity range $\pm 350$ km s$^{-1}$.

This represents the best-constrained relation at this redshift range available in the literature so far, in terms of the number of coadded spectra and mass bins.

### 4.2. Comparison with Other Literature Results

Figure 7 shows a comparison between our scaling relations and others in the literature at a similar redshift. Specifically, we report the relation obtained solely from MIGHTEE-H I at $z \sim 0.36$ (S22) and a recent GMRT result at $z \sim 0.35$ (A. Bera et al. 2023a).

Our scaling relation presents significant differences from that reported in S22. This is discussed in greater detail in Appendix B. In summary, we consider our scaling relation as an improved updated reference at this redshift with respect to the result from S22, as a result of an augmented catalog obtained with a more conservative redshift quality cut, as well as larger statistics, a more accurate selection of star-forming galaxies, and the implementation of an RFI masking technique. We also find tension between our scaling relation and the relation extracted by A. Bera et al. (2023a) at $z \sim 0.35$, which lies slightly below those in the local Universe (Figure 8), supporting the hypothesis of star-forming galaxies with $\log M_\star/M_\odot > 9.5$ at $z \sim 0.35$ being H I-poorer than those at $z \sim 0$. We suggest that one of the drivers of this dissimilarity might be the different selection criterion for star-forming galaxies. While A. Bera et al. (2023a) select star-forming galaxies based on a $U - B$ color, our two-color criterion is designed to be less sensitive to reddening and therefore should better capture dusty red star-forming galaxies and remove quenched objects (as discussed in Section 2.1.2). Moreover, the cosmic volumes probed by A. Bera et al. (2023a) differ significantly from ours. The scaling relation obtained in our work is extracted from a cosmic volume ($\approx 10^6$ cMpc$^3$) that is $\sim 18$ times larger than A. Bera et al. (2023a) and is based on a larger spectroscopic sample of stacked sources (approximately by a factor of 16).

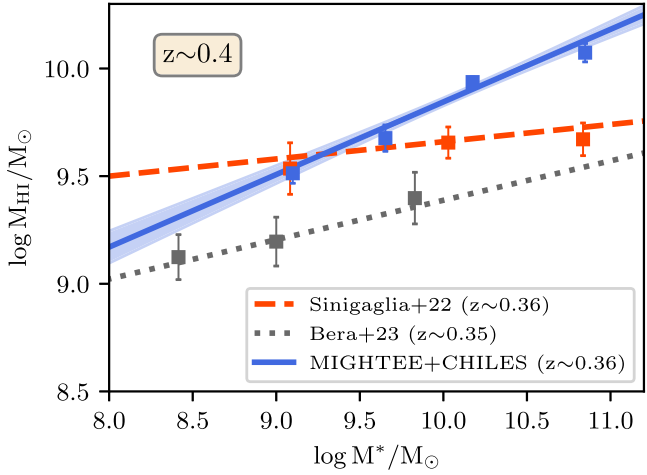

**Figure 7.** $M_{H\,I}$–$M_\star$ scaling relation for star-forming galaxies at $z \sim 0.36$. Our stacking results are displayed in blue squares, fitted by the blue line with the relative uncertainty (blue shaded area). The orange squares represent the current reference relation at $z \sim 0.36$ from MIGHTEE-H I (S22), alongside the relative fit (orange dashed line). Gray squares and the gray dotted line display the relation extracted with GMRT data at $z \sim 0.35$ by A. Bera et al. (2023a).

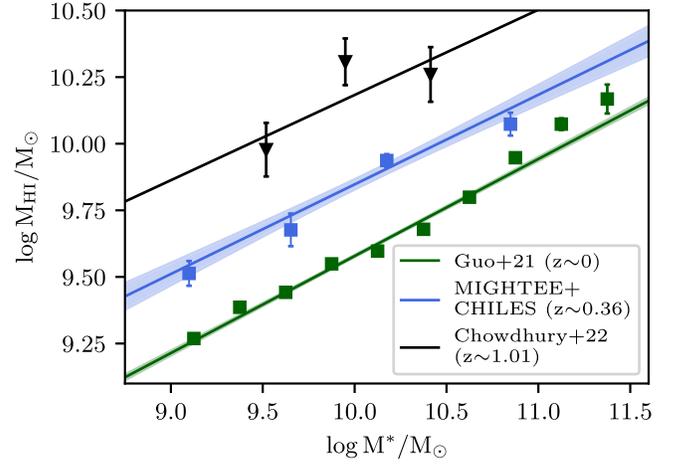

**Figure 9.** Evolution of the $M_{H\,I}$–$M_\star$ scaling relation for star-forming galaxies. Our stacking results ($z \sim 0.36$) are displayed as blue squares, fitted by the blue linear law with the related uncertainty (blue shaded area). Green squares and the green solid line represent the stacked points and the corresponding fit for star-forming galaxies extracted from G21 at $z \sim 0$, and black triangles and the black solid line are taken from C22 at $z \sim 1$.

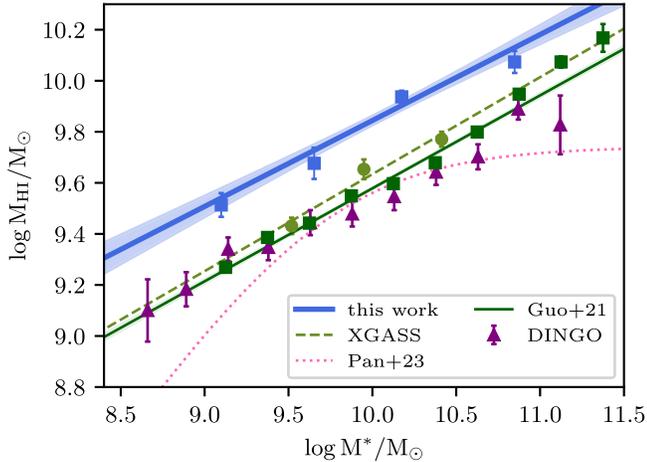

**Figure 8.** $M_{H\,I}$–$M_\star$ scaling relations measured from star-forming galaxies in the nearby Universe. The data points and best-fit curves—when both are available—are represented with the same color coding.

We also compare our result with other scaling relations in the nearby Universe. Early results from HIPASS (C. Evoli et al. 2011) report a $\log(M_{HI}/M_\star) - \log M_\star$ relation based on kinematically inferred HI masses from galaxies in the local Universe, with a slope of approximately –0.5, in great agreement with predictions by numerical models (A. B. Romeo et al. 2020). That corresponds to a 0.5 slope in the $\log M_{HI} - \log M_\star$ plane, consistent with the slopes measured in the local Universe displayed in Table 2. In Figure 8, we plot our scaling relation together with the one by G21, obtained from spectral stacking of star-forming galaxies from the ALFALFA (R. Giovanelli et al. 2005; M. P. Haynes et al. 2018) survey cross-matched with the SDSS (D. G. York et al. 2000) DR7 Main Galaxy Sample. We also display the scaling relation based on MIGHTEE-H I data extracted at $z \sim 0$

(H. Pan et al. 2023). Moreover, we include results from the DINGO survey (J. Rhee et al. 2023), also based on stacking, and from C22, who provided a scaling relation at $z \sim 0$ based on a reshuffling of the blue xGASS galaxies (B. Catinella et al. 2018). We highlight that the scaling relation that we measure in this work at $z \sim 0.36$ is not consistent with any of the results at lower redshift and has systematically larger normalization. This supports the argument that the H I content of star-forming galaxies undergoes a nonnegligible evolution with redshift.

### 4.3. Evolution of the $M_{H\,I}$–$M_\star$ Scaling Relation

To visualize the evolution of the $M_{H\,I}$–$M_\star$ scaling relation, we plot in Figure 9 the local relation by G21 along with our results at $\langle z \rangle = 0.36$ and the results from C22 at $\langle z \rangle = 1$, also obtained from star-forming galaxies. We summarize the best-fitting parameters (slope and offset) of some of the aforementioned scaling relations with their reference redshift in Table 2. We also report the $\log M_{HI}$ predicted at $\log(M_\star/M_\odot) \sim 10$ (the bulk of the stellar mass distribution of our sample) by such relations to quantify its evolution with redshift. For consistency of notation with C22, we reduced the equation to the form $\log(M_{HI}/M_\odot) = \alpha \log(M_\star/M_\odot) + \beta$. Figure 9 and Table 2 show that the slope does not seem to undergo a strong evolution with redshift. In contrast, there is evidence for evolution in terms of normalization. In particular, for a fixed stellar mass, our scaling relation at $z \sim 0.36$ yields $\sim 60\%$ more atomic mass than its counterpart in the local Universe and $\sim 50\%$ less atomic mass than what is predicted by C22 at $z \sim 1$.

The increase in normalization of $M_{H\,I}$ can be quantified more accurately by looking at the evolution of the atomic hydrogen mass content with redshift for a fixed stellar mass. Figure 10 shows the $\log M_{HI} - \log(1 + z)$ dependence for $\log(M_\star/M_\odot) = 10$. We include two data points at $z = 0$ and $z = 1$ extracted from the two scaling relations reported by G21 and C22 based on ALFALFA and high-redshift H I GMRT data, respectively. Additionally, we incorporate one point extracted from our scaling relation at $z = 0.36$. We fit the three data points with a power-law function in an attempt to quantify

Table 2
Slope and Offset of Scaling Relations ($\log(M_{HI}/M_\odot) = \alpha \log(M_\star/M_\odot) + \beta$) Extracted from Different Surveys and Redshift Bins

| Survey | Redshift | $\alpha$ | $\beta$ | $M_{HI}(\log M_\star \sim 10)$ | References |
|---|---|---|---|---|---|
| ALFALFA | $z = 0.0025$–$0.06$ | 0.42 | 5.35 | 9.55 | G21 |
| xGASS | $z = 0.01$–$005$ | $0.38 \pm 0.05$ | $5.83 \pm 0.20$ | $9.63 \pm 0.05$ | B. Catinella et al. (2018); C22 |
| MIGHTEE+CHILES | $\langle z \rangle = 0.36$ | $0.32 \pm 0.04$ | $6.65 \pm 0.36$ | $9.83 \pm 0.03$ | This work |
| CATz1 | $\langle z \rangle = 1.01$ | $0.32 \pm 0.13$ | $6.98 \pm 0.57$ | $10.09 \pm 0.12$ | C22 |

**Note.** We report the reference survey (first column), redshift range or median redshift (second column), slope and offset (third and fourth columns), predicted hydrogen mass at $\log(M_\star/M_\odot) \sim 10$ (fifth column), and reference publications (last column). We list the aforementioned results with increasing redshift from top to bottom.

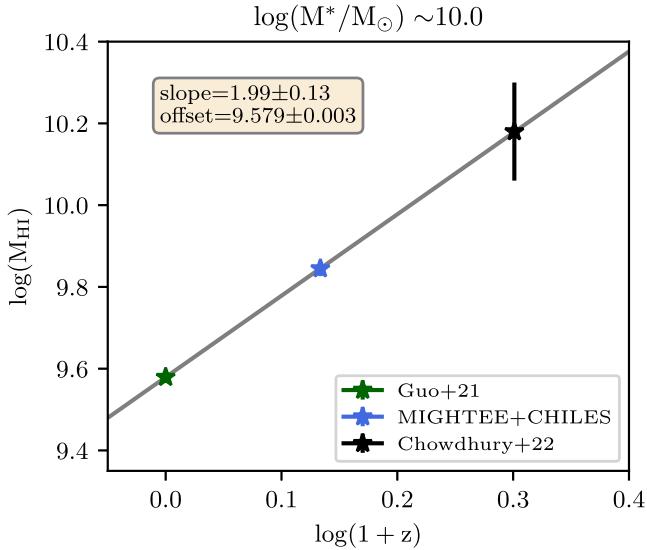

**Figure 10.** Atomic hydrogen mass evolution as a function of redshift. We display the points extracted at $\log(M_\star/M_\odot) \sim 10$ for three different scaling relations, at $z = 0$ from our fit to ALFALFA data (G21), $z = 0.36$ from the combined MIGHTEE+CHILES stack, and $z = 1$ from CATz1 (C22). The gray line shows a simple linear fit, where slope and offset are shown in the yellow box (top left corner).

the evolution of the hydrogen mass content with redshift. We obtain a power-law index of $1.99 \pm 0.13$.

To have a further element of comparison with these observational results, we use the output of NEUTRALUNIVERSEMACHINE (NUM; H. Guo et al. 2023). NUM is an empirical model that parameterizes the probability distribution of SFRs in dark matter halos as a function of several parameters, including redshift, maximum halo circular velocity at the redshift of the peak halo mass ($\nu_{M_{peak}}$), and relative change in $\nu_{max}$ over the past halo dynamical time ($\Delta\nu_{max}$). The latter two parameters are used as a proxy for the mass accretion rate, while the SFR is drawn from some probability distributions. The stellar mass is obtained by integrating the SFR. Once these values are fixed, all others are determined by fitting the available observations in the NUM catalog to stellar mass functions, cosmic SFRs, sSFRs, quenched fractions, UV luminosity functions, UV–stellar mass relations, and correlation functions. We refer the reader to P. Behroozi et al. (2019) for more details. NUM predicts the mass of the atomic hydrogen in a given galaxy halo as a function of its virial mass, redshift at the time of formation, SFR, and redshift. We compare the predictions of the model with our main observational data in Figure 11. First, we assume the scaling relation at $z \sim 0$ by G21 and scale it to higher redshift by using the power-law index $1.99 \pm 0.13$ derived from Figure 10 at fixed stellar mass, assuming a constant slope and shifting the normalization to higher H I masses. We plot the results of this procedure at $z = 0$ (original fitting function by G21; green solid), $z = 0.36$ (blue dotted), and $z = 1.01$ (black dotted) in the left panel of Figure 11. In the same panel, we also plot again the observational constraints from G21 for the local Universe, at $z \sim 0.36$ from this work and $z \sim 1$ from C22. As expected, the rescaled relation provides a good match with stacking observations at higher redshifts. NUM snapshots of the scaling relation at $z = 0$, $z = 0.36$, and $z = 1.12$ are shown in the right panel of Figure 11 (solid lines) with the same observational references as in the left panel. We note that, while underpredicting $M_{H I}$, NUM maintains a somewhat constant slope with stellar mass. Also, the evolution with redshift predicted by NUM is qualitatively in agreement with the one found from observational data but quantitatively shallower. In fact, by fitting the H I masses obtained from the three NUM redshift snapshots at $\log M_\star/M_\odot \sim 10$ in the same fashion as we did for the observational relations, we obtain an evolutionary index of $\sim 1.6$, roughly 20% lower than what is predicted by observations at the same stellar mass. This might be connected with the fact that NUM is tuned to fit the H I content of galaxies in the nearby Universe, such as ALFALFA, xGASS, and xCOLD GASS (A. Saintonge et al. 2017). See H. Guo et al. (2023) for more details.

## 5. Discussion

We measured the $M_{H I}$–$M_\star$ scaling relation at $z \sim 0.36$ by combining spectra from the MIGHTEE-H I survey and CHILES. In the following, we discuss the main implications of our results. We start by clarifying that the neutral hydrogen mass we are coadding with our stacking process is integrated over the full extent of the galaxies. In fact, by adopting a $3\times$ beam aperture, we aim to capture the amount of atomic gas both inside and outside the optical disk (H I reservoirs). As a result, the scaling relations we derive offer a global measurement of the H I content of galaxies, rather than providing spatially resolved information about the relative amount of atomic gas found inside or outside the optical disk.

We place our scaling relation at $\langle z \rangle = 0.36$ within the framework of cosmic evolution by comparing it to the results at $z \sim 0$ and $z \sim 1$. Assuming a linear model for the scaling relation in logarithmic scale in this redshift range, we argue that there is little evolution in terms of slope ($<1\sigma$). In this redshift range, galaxies have depleted their H I reservoirs with minimal dependence on their stellar mass. This suggests that over the last 8 Gyr, the physical processes that govern H I replenishment and depletion are similar for galaxies with different stellar masses. A natural mechanism accounting for such a scenario is represented by cold gas accretion (e.g., C. J. Conselice et al. 2013; S. Cantalupo et al. 2014; E. Daddi et al. 2022), with a potential contribution from galaxy mergers (at least at $z < 0.5$; e.g., J. Sánchez Almeida et al. 2014; H. Padmanabhan & A. Loeb 2020). In particular, we have quantified the growth of H I at $\log M_\star \sim 10$ with redshift and found $M_{\rm H\,I} \propto (1 + z)^{1.99}$. For comparison, Equation (15) in L. Morselli et al. (2021) reports a power-law $M_{\rm H\,I}/M_\star \propto (1 + z)^{2.9}$. However, we caution that the work by L. Morselli et al. (2021) extrapolates the global $M_{\rm H\,I}/M_\star$–$z$ evolution from internal (spatially resolved) relations in nearby galaxies.

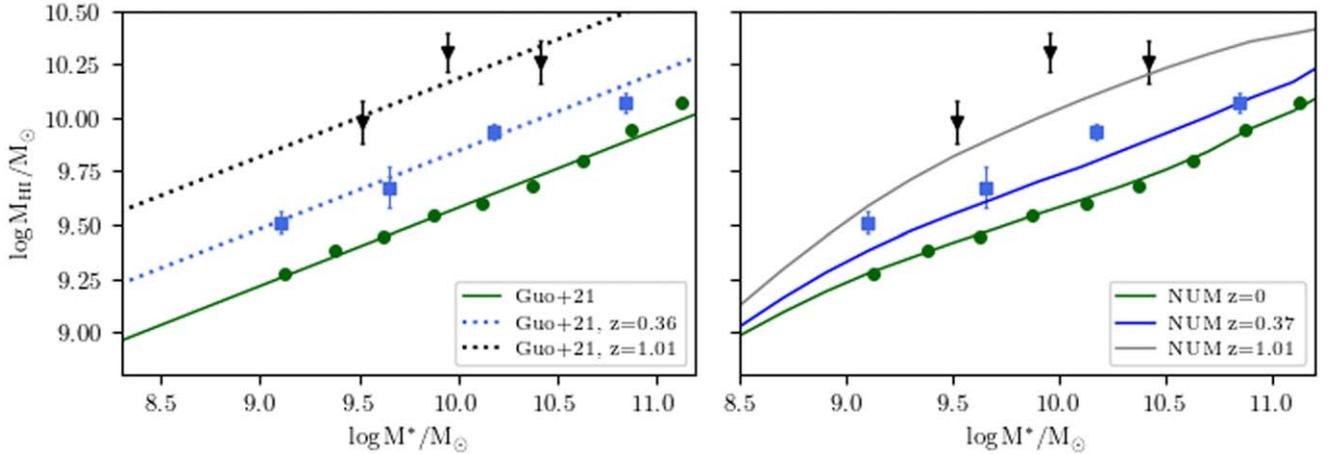

**Figure 11.** Evolution of the $M_{\rm H\,I}$–$M_\star$ scaling relation for star-forming galaxies. Left panel: we plot the scaling relation at $z \sim 0$ computed by G21 and upscale it with the evolutionary power-law index 1.8 at redshift $z = 0.36$ and $z = 1.01$. We also add observational data: data points from G21 (green circles), combined MIGHTEE–CHILES results ($z \sim 0.36$; blue squares) and $z \sim 1$ data from C22 (black squares). Right panel: curves from three different redshift snapshots ($z = 0$, $z = 0.36$, and $z = 1.12$) are displayed from NUM (H. Guo et al. 2023). Observational data are color-coded as in the left panel.

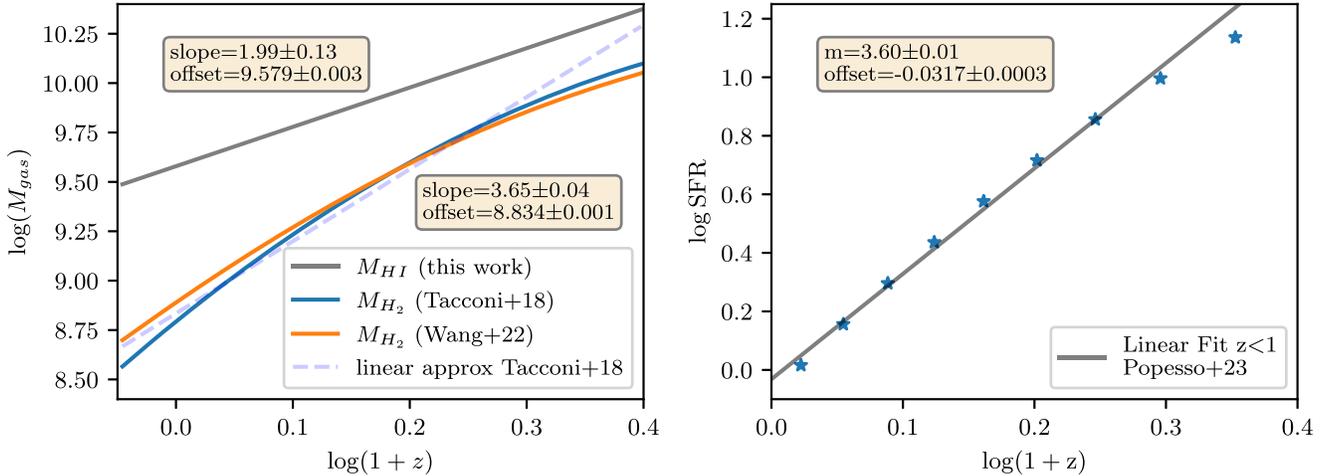

**Figure 12.** A comparison between the estimated, first-order evolutionary trend of gas and SFR in the range $0 < z < 1$ and at fixed stellar mass ($\log M_\star = 10$). Left panel: atomic hydrogen mass as a function of redshift (solid gray line), compared to the molecular gas model (L. J. Tacconi et al. 2018; solid blue line), approximated by a linear fit in this range (dashed light blue line), and T.-M. Wang et al. (2022; solid orange line). The slopes of the two linear relations are shown in the yellow text boxes next to the relation to which they refer. Right panel: SFR extracted at $\log M_\star = 10$ from the evolving MS predicted in P. Popesso et al. (2023). A linear fit with redshift is performed, and the resulting parameters are shown in the yellow text box.

The H I evolution measured in this paper could be compared with the reported redshift evolution of SFR and molecular gas to study the dynamics of the baryon cycle as a function of cosmic time. Evolutionary studies on the MS from the literature report that it monotonically increases its normalization with $z$ (for a fixed stellar mass). A. L. Faisst et al. (2016) quantified this evolution in terms of the linear increase in sSFR (defined as the SFR per unit of stellar mass, sSFR = SFR/$M_\star$) with $\log(1 + z)$, finding an evolutionary index of $\sim$2.4 at $z < 2.2$. Other results exploring the log sSFR – log$(1 + z)$ correlation indicate different trends: J. S. Speagle et al. (2014) report an evolutionary index of $\sim$2.8, while S. Oliver et al. (2010) and A. Karim et al. (2011) find a coefficient of $\sim$3.5 at $z < 2.5$. We can also take the MS as modeled by P. Popesso et al. (2023) at $0 < z < 1$ and derive the evolution of the SFR with $z$ for a fixed stellar mass (e.g., $\log M_\star/M_\odot \sim 10$). This exercise is presented in Figure 12 (right panel), where we also report a linear fit to the log(SFR) – log$(1 + z)$ relation up to $z \sim 1$, obtaining a slope of $\sim$3.6.

The SFR and the $H_2$ content in galaxies are expected to tightly correlate, the latter being the fuel for star formation activity. In the left panel of Figure 12, we plot the trend of the evolving content of molecular gas $M_{H_2}$ for a fixed stellar mass $\log M_\star/M_\odot = 10$, as modeled by L. J. Tacconi et al. (2018):

$$\log \frac{M_{H_2}}{M_\star} = 0.07 - 3.8 \cdot [\log(1+z) - 0.63]^2 \\ - 0.33 \cdot (\log M_\star - 10.7). \quad (5)$$

When we approximate this law, which is quadratic in $\log(1+z)$, with a linear relation in the $0 < z < 1$ range, we obtain a slope of ~3.6. This coefficient is remarkably similar to the one we estimated for the evolution of the MS based on P. Popesso et al. (2023). As expected, this means that the rate of star formation and the content of molecular gas evolve at similar rates, suggesting that molecular depletion time is not a strong function of redshift. T.-M. Wang et al. (2022) used a similar functional form to fit the evolution of $M_{H_2}$: in that case, the approximated first-order slope is roughly 3.2.

Overall, we find that the rate of the evolution of H I content in star-forming galaxies seems to be shallower with respect to the rates of $H_2$ consumption and SFR evolution. The fact that the amount of atomic gas decreases at a slower rate than that of molecular gas (see also F. Walter et al. 2020) suggests that the baryon cycle may experience a bottleneck in the H I → $H_2$ conversion (F. Bigiel et al. 2010; S22; A. Bera et al. 2023b). In this sense, recent results from MHONGOOSE (W. J. G. de Blok et al. 2024) and from the PHANGS-MeerKat collaboration find evidence for a low inward radial mass flow rate in the disk (S. Laudage et al. 2024), supporting the probable lack of fuel for star formation in the inner regions of galaxies, and discuss the suitable conditions for the conversion from the atomic to the molecular phase (C. Eibensteiner et al. 2024). The reason for this might be related to H I being unable to enter the star-forming disk to sustain star formation, as it managed to do at cosmic noon ($z \sim 2$). At lower redshift, we speculate that galaxy angular momentum plays an important role in keeping neutral hydrogen in the outer H I disk, preventing it from accessing the optical disk of the galaxy, where star formation takes place (Y.-j. Peng & A. Renzini 2020). In support of this hypothesis, empirical and numerical evidence shows a tight correlation between gas fraction and baryonic angular momentum, such that high angular momentum disk galaxies are also more gas-rich and most of the angular momentum is contained into atomic gas (P. E. Mancera Piña et al. 2021a, 2021b; A. B. Romeo et al. 2023). As a final note, we shall bear in mind that we are studying extended H I profiles, while the trends we discussed for SFR and $H_2$ are mostly confined to the inner part of a galaxy (i.e., roughly corresponding to the optical radius). Higher-resolution studies with the SKA will allow us to carry out a more significant comparison at the same physical scales. A complete physical interpretation of the processes competing in the baryon cycle will be investigated further with the support of hydrodynamical simulations—such as, e.g., SIMBA (L. J. M. Davies et al. 2018), IllustrisTNG (D. Nelson et al. 2019), and EAGLE (J. Schaye et al. 2015)—in a future paper.

The SuperMIGHTEE survey (D. Lal et al. 2025, in preparation) represents a suitable data set to add information at higher redshifts. It covers the COSMOS field using uGMRT Band 4 ($\approx 550-850$ MHz) and the XMM-LSS field using Band 3 ($\approx 250-500$ MHz) and Band 4 (Y. Gupta et al. 2017), potentially pushing the scaling relation to $z \gtrsim 1$.

## 6. Summary

We summarize the main findings of this paper here.

1. We presented a novel global $M_{H\,I}$–$M_\star$ scaling relation at redshift $z \sim 0.36$ by stacking spectra from the MIGHTEE-H I survey and CHILES together, achieving the most robust statistics for a stacking project at this redshift to date. The best-fit power-law model to our data is $\log_{10} M_{HI} = (0.32 \pm 0.04) \times \log_{10} M_\star + (6.65 \pm 0.36)$.
2. When comparing the scaling relation at $z \sim 0.36$ with its counterpart at $z \sim 0$ and $z \sim 1$, we observe an evolution only in terms of normalization, while the slope does not show evidence of significant evolution. This suggests that the same stellar-mass-independent processes have governed H I replenishment and depletion over the last 8 Gyr.
3. We quantified the evolution of the scaling relation with redshift. In particular, at fixed mass $\log M_\star \sim 10$, we find the following best-fitting relation: $\log M_{HI} = (1.99 \pm 0.13) \cdot \log(1+z) + (9.579 \pm 0.003)$. We noticed that this evolutionary power-law function can be related to the one linking $\log M_{H_2}$ and $\log$ SFR (MS) with $\log(1+z)$. The latter two trends can be linearly approximated with the same slope, roughly 3.6, suggesting balanced rates of star formation and molecular gas consumption. The fact that the coefficient for $\log M_{HI}$ is smaller might indicate that neutral gas reservoirs decrease at a slower rate than molecular gas, hinting at a bottleneck in the H I → $H_2$ conversion in the context of the baryon cycle. We speculate that this might be connected with the fact that the angular momentum of the galaxy may be keeping neutral hydrogen in the outer H I disk, preventing it from accessing the optical disk where star formation occurs. Further investigation with hydrodynamical simulations is planned for a future paper.

## Acknowledgments

A.B. acknowledges the support of the doctoral grant funded by the University of Padova and by the Italian Ministry of Education, University and Research (MIUR). This work was supported in part by the Italian Ministry of Foreign Affairs and International Cooperation, grant No. ZA23GR03. F.S. acknowledges the support of the Swiss National Science Foundation (SNSF) 200021_214990/1 grant. G.R. acknowledges the support from grant PRIN MIUR 2017-20173ML3WW 001. A.B., F.S., G.R., L.B., and I.P. acknowledge support from INAF under the Large Grant 2022 funding scheme (project "MeerKAT and LOFAR Team up: a Unique Radio Window on Galaxy/AGN co-Evolution"). A.B. acknowledges the help of Prof. Hong Guo and Dr. Jonghwan Rhee for providing material from their own research work and also Prof. J.H. Van Gorkom for providing in-depth comments on this manuscript. I.P. and A.B. acknowledge financial support from the South African Department of Science and Innovation's National Research Foundation under the ISARP RADIOMAP Joint Research Scheme (DSI-NRF grant No. 150551) and from the Italian Ministry of Foreign Affairs and International Cooperation under the "Progetti di Grande Rilevanza" scheme (project RADIOMAP, grant No. ZA23GR03). M.V. acknowledges financial support from the Inter-University Institute for Data Intensive Astronomy (IDIA), a partnership of the University of Cape Town, the University of Pretoria, and the University of the Western Cape, and from the South African Department of Science and Innovation's National Research Foundation under the ISARP RADIO-SKY2020 and RADIOMAP Joint Research Schemes (DSI-NRF grant Nos. 113121 and 150551) and the SRUG HIPPO Projects (DSI-NRF grant Nos. 121291 and SRUG22031677). D.J.P. greatly acknowledges support from the South African Research Chairs Initiative of the Department of Science and Technology and National Research Foundation. M.B. acknowledges financial support from the Flemish Fund for Scientific Research (FWO-Vlaanderen, Bilateral Scientific Cooperation grant No. G0G0420N) and the Belgian Science Policy Office (BELSPO Cooperation Grant GALSIMAS, BL/02/SA12). The NRAO is a facility of the National Science Foundation operated under cooperative agreement by Associated Universities, Inc. The MeerKAT telescope is operated by the South African Radio Astronomy Observatory, which is a facility of the National Research Foundation, an agency of the Department of Science and Innovation. This work made use of the IDIA processMeerKAT pipeline, developed at the Inter-University Institute for Data Intensive Astronomy (IDIA), available at https://idia-pipelines.github.io (DOI:10.23919/URSIGASS51995.2021.9560276). We acknowledge the use of the ilifu cloud computing facility, www.ilifu.ac.za, a partnership between the University of Cape Town, the University of the Western Cape, the University of Stellenbosch, Sol Plaatje University, the Cape Peninsula University of Technology, and the South African Radio Astronomy Observatory. The Ilifu facility is supported by contributions from the Inter-University Institute for Data Intensive Astronomy (IDIA—a partnership between the University of Cape Town, the University of Pretoria, the University of the Western Cape, and the South African Radio Astronomy Observatory), the Computational Biology division at UCT, and the Data Intensive Research Initiative of South Africa (DIRISA). This work made use of the CARTA software (Cube Analysis and Rendering Tool for Astronomy; A. Comrie et al. 2021; DOI:10.5281/zenodo.3377984, https://cartavis.github.io). This research used data obtained with the Dark Energy Spectroscopic Instrument (DESI). DESI construction and operations are managed by the Lawrence Berkeley National Laboratory. This material is based upon work supported by the U.S. Department of Energy, Office of Science, Office of High-Energy Physics, under contract No. DE-AC02-05CH11231, and by the National Energy Research Scientific Computing Center, a DOE Office of Science User Facility under the same contract. Additional support for DESI was provided by the U.S. National Science Foundation (NSF), Division of Astronomical Sciences, under contract No. AST-0950945 to the NSF's National Optical-Infrared Astronomy Research Laboratory; the Science and Technologies Facilities Council of the United Kingdom; the Gordon and Betty Moore Foundation; the Heising-Simons Foundation; the French Alternative Energies and Atomic Energy Commission (CEA); the National Council of Science and Technology of Mexico (CONACYT); the Ministry of Science and Innovation of Spain (MICINN); and the DESI Member Institutions: www.desi.lbl.gov/collaborating-institutions. The DESI collaboration is honored to be permitted to conduct scientific research on Iolkam Du'ag (Kitt Peak), a mountain with particular significance to the Tohono O'odham Nation. Any opinions, findings, and conclusions or recommendations expressed in this material are those of the author(s) and do not necessarily reflect the views of the U.S. National Science Foundation, the U.S. Department of Energy, or any of the listed funding agencies. DEVILS is an Australian project based around a spectroscopic campaign using the AAT. The DEVILS input catalog is generated from data taken as part of the ESO VISTA-VIDEO (M. J. Jarvis et al. 2013) and UltraVISTA (H. J. McCracken et al. 2012) surveys. DEVILS is partly funded via Discovery Programs by the Australian Research Council and the participating institutions. The DEVILS website is https://devilsurvey.org. The DEVILS data are hosted and provided by AAO Data Central (https://datacentral.org.au).

## Appendix A
## Separating CHILES and MIGHTEE

Testing properties and comparing individual results for MIGHTEE and CHILES is fundamental to see whether or not the two surveys can be combined together.

### A.1. Gaussianity Tests

Radio data reduction is a complex procedure, and it includes several steps, such as flagging, calibration, and image reconstruction. Throughout this process, it is common to assume Gaussianity of the noise, but any sources of inaccuracies propagate over data reduction and may result in a certain degree of non-Gaussian behavior. In fact, the assumption of Gaussianity can be easily broken in the aforementioned steps, for example, due to bad flagging, intruder noncalibrator sources, or RFI. Before stacking, we test the Gaussianity of the set of spectra that will enter the selection. Figure 5 shows that the noise scales properly with the number of randomly placed stacked spectra, as expected from Gaussian statistics (rms $\propto \sqrt{N}$). Moreover, we consider the full set of spectra to be stacked from MIGHTEE and CHILES and compute the mean of the fluxes of their channels, excluding the channels that fall within the integration range ($\pm 350 \, \text{km s}^{-1}$). We then show the distributions of the means for both surveys, fitted by a Gaussian model, in Figure 13. A Kolmogorov–Smirnov (KS) test between the fit and the model was carried out for both surveys. Such a test turned positive for the MIGHTEE distribution, rejecting the hypothesis that the data were drawn from a Gaussian distribution with a $p$-value of 0.05. At this point, we looked for a conservative strategy to ensure Gaussianity. Specifically, the estimated $\sigma$ parameter of the distribution provides a useful criterion to rule out outliers, i.e., excessively noisy spectra. We choose $3\sigma$ as an upper limit for the selection of spectra for stacking. After the cut, the KS test between model and distribution did not reject the compatibility with a Gaussian model in both cases.

### A.2. CHILES versus MIGHTEE Stacking Comparison

We checked the consistency of the H I masses obtained by stacking the spectra of galaxies in the same stellar mass bins from the two data cubes separately. This is to test if cosmic variance between the two survey footprints is causing inconsistencies on the estimated H I masses. We split the full sample into three stellar mass bins instead of four, as we did for the main result of the paper, to ensure that we have enough statistics when stacking the two surveys separately: $8.0 < \log M_\star < 9.5$, $9.5 < \log M_\star < 10.5$, and $\log M_\star > 10.5$. Figure 14 shows the stacked spectra for MIGHTEE and CHILES in the three different stellar mass bins after having excluded spectra falling within a spectral interval affected by the strong RFI mask (see Section 3) or with mean flux beyond the threshold (see Appendix A.1). The $M_{\rm H\,I}$ estimates and their uncertainty, as well as the number of stacked galaxies and the resulting S/N, are reported inside the panels. Even though CHILES appears to be collecting systematically less mass than MIGHTEE, the measurements are consistent within $1.5\sigma$ in all the probed stellar mass bins. Figure 15 instead shows the resulting measurements in the $M_{\rm H\,I}$–$M_\star$ plane, displaying those of MIGHTEE as blue stars and those of CHILES as brown triangles. We also fit the masses separately for the two surveys, using a linear model and obtaining two sets of fitting parameters that are consistent with each other and are reported in Equations (A1) and (A2) for CHILES and MIGHTEE, respectively:

$$\log_{10} M_{\rm HI} = (0.33 \pm 0.06)\log_{10} M_\star + (6.55 \pm 0.59), \quad (A1)$$

$$\log_{10} M_{\rm HI} = (0.29 \pm 0.03)\log_{10} M_\star + (7.03 \pm 0.30). \quad (A2)$$

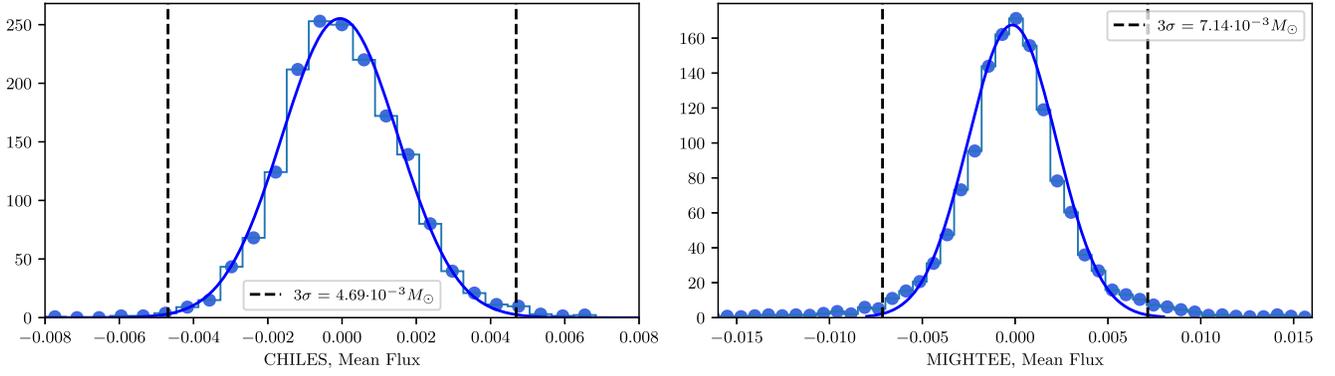

Figure 13. Distribution of mean channel fluxes (Jy beam$^{-1}$) per each stacked spectrum in MIGHTEE, excluding the channels lying within the integration region.

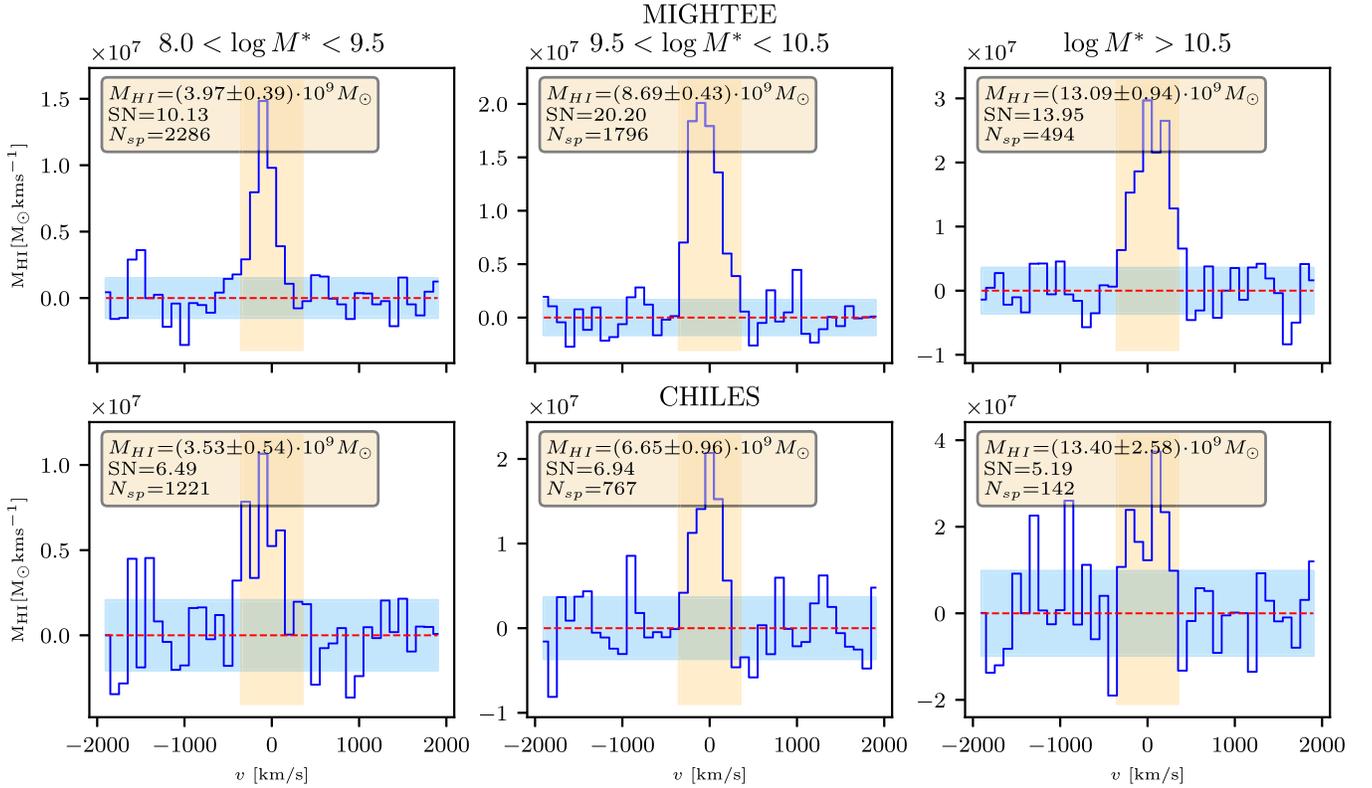

Figure 14. A direct comparison of the resulting stacked spectra from MIGHTEE (top row) and CHILES (bottom row) over the same stellar mass bins. We report the stacked spectra as blue solid lines. The bin width used is $\Delta v = 100$ km s$^{-1}$. The text box in each panel reports the H I mass estimate with the associated uncertainty, as well as the S/N and the total number of stacked galaxies. The light blue shaded area highlights the noise level assigned to the stacked spectrum, computed as the rms of the flux densities in the bins outside the integration range.

This highlights the consistency between the two separate scaling relations extracted from two surveys conducted with different instruments and supports the reliability of the combined scaling relation presented in this paper.

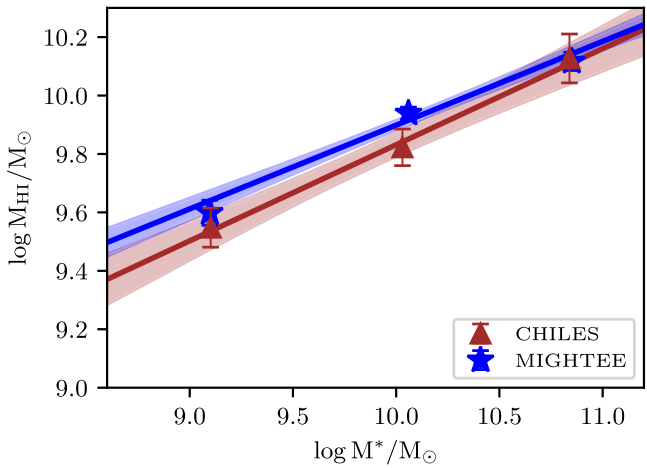

**Figure 15.** A comparison between the $M_{HI}$–$M_\star$ scaling relation obtained from MIGHTEE (blue stars) and CHILES (brown triangles) and the best-fitting linear relations to the measurements for each survey (solid lines).

## Appendix B
## Inconsistency with S22

We note that the stacking on the Early Science MIGHTEE data cubes was already performed in S22, but we update it here in light of methodological changes (e.g., implementation of RFI masking) and the improved input catalog. We point out that the H I masses that we measure in this work from MIGHTEE data alone are not consistent with the scaling relation presented in S22. Although the measured mass of H I in the lowest mass bin is in excellent agreement with the analogous measurement from S22, the two measurements at higher $M_\star$ systematically yield higher $M_{HI}$. To fully understand the origin of this inconsistency, we analyze the various differences between the two analyses in what follows.

1. *Difference in the photometric catalog used.* Although here we select our sample based on the same criteria as in S22 (see Section 2.1), we use the photometry from the COSMOS2020 catalog (J. R. Weaver et al. 2022) instead of COSMOS2015 (C. Laigle et al. 2016).
2. *Difference in the spectroscopic catalog used.* The spectroscopic catalog adopted in this work (A. Khostovan et al. 2025, in preparation) consists of an updated version of the one used in S22. In particular, it includes additional spectroscopic sources from DESI and DEVILS, as well as a refined prioritization of the reliability of spectroscopic redshift, when multiple redshifts for the same source are available from different surveys. Finally, we introduced a quality cut on spectroscopic redshift, as explained in Section 2.1.1, which was not featured in S22.
3. *Introduction of the RFI flagging strategy.* This has already been discussed and presented as a novel methodological aspect in Section 3.

To assess which of these aspects is the major driver of the change in the slope of the MIGHTEE scaling relation, we perform a series of tests. We start by testing the impact of the catalog. In particular, Figure 16 displays a comparison between the following setups: (i) old catalog without RFI masking (i.e., same as S22; magenta triangles), (ii) old catalog with RFI masking (green circles), (iii) new catalog without RFI masking (brown stars), (iv) new catalog with RFI masking (blue squares), and (v) creating a catalog by matching the photometry from COSMOS2015 with the updated spectroscopic catalog (black circles). Also, to make the comparison more straightforward, we use the same three stellar mass bins as in S22 (i.e., no lower mass cut at $\log M_\star$). This figure highlights that the RFI flagging strategy does not introduce systematic biases, while the change in catalog appears to have the greatest impact on the results. In particular, the COSMOS2020 photometry features tiny deviations from the COSMOS2015 one, and we have explicitly verified that such differences have a negligible impact on the stacking results (black circles). In contrast, the change in the scaling relation is rather due to the differences in the spectroscopic redshift between the two catalogs. In fact, redshift uncertainties are on the order of $\sim 5 \times 10^{-4}$, which translates into a typical velocity uncertainty of $\sim 150 \, \mathrm{km \, s^{-1}}$ in the position of the zero-velocity channel of the spectrum. We conclude that this aspect can have a fairly strong impact on the stacking procedure. Moreover, this effect might be stronger as we go to the high-mass end, where there are fewer galaxies and redshift inaccuracy is more likely to cause a loss in the collected H I mass. This highlights the importance of a proper redshift quality cut.

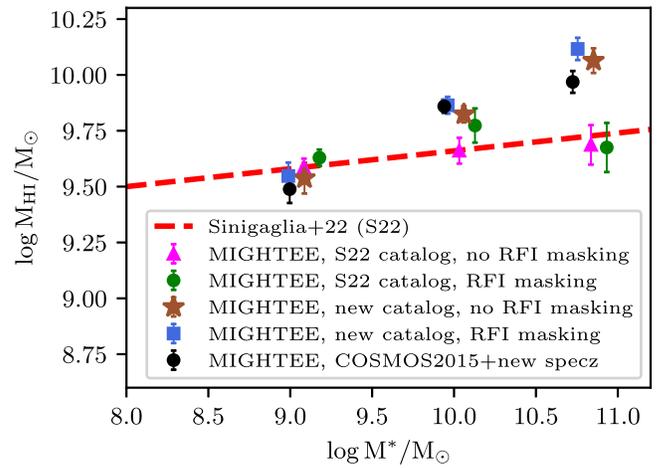

**Figure 16.** A comparison between the scaling relations obtained herein from MIGHTEE data with different setups and from S22. The magenta triangles are the estimated H I masses obtained by running the pipeline with the same setup as in S22, i.e., without RFI masking and using the previous version of the spectroscopic catalog. The red dashed line is the best fit presented in S22. Green circles are the estimated H I masses from the same catalog as in S22 but including the RFI flagging strategy. Cyan squares and brown stars refer to the new catalog used in this work, with and without RFI masking, respectively. An artificial negative/positive offset was added along the x-axis to the cyan squares and green and black circles to ease the visual comparison.

### ORCID iDs

Alessandro Bianchetti ⓘ https://orcid.org/0009-0002-8916-3430
Francesco Sinigaglia ⓘ https://orcid.org/0000-0002-0639-8043
Giulia Rodighiero ⓘ https://orcid.org/0000-0002-9415-2296
Ed Elson ⓘ https://orcid.org/0000-0001-9359-0713
Mattia Vaccari ⓘ https://orcid.org/0000-0002-6748-0577
D.J. Pisano ⓘ https://orcid.org/0000-0001-7996-7860
Isabella Prandoni ⓘ https://orcid.org/0000-0001-9680-7092
Kelley Hess ⓘ https://orcid.org/0000-0001-9662-9089
Maarten Baes ⓘ https://orcid.org/0000-0002-3930-2757
Elizabeth A.K. Adams ⓘ https://orcid.org/0000-0002-9798-5111

Filippo M. Maccagni https://orcid.org/0000-0002-9930-1844
Alvio Renzini https://orcid.org/0000-0002-7093-7355
Laura Bisigello https://orcid.org/0000-0003-0492-4924
Min Yun https://orcid.org/0000-0001-7095-7543
Emmanuel Momjian https://orcid.org/0000-0003-3168-5922
Hansung B. Gim https://orcid.org/0000-0003-1436-7658
Hengxing Pan https://orcid.org/0000-0002-9160-391X
Thomas A. Oosterloo https://orcid.org/0000-0002-0616-6971
Richard Dodson https://orcid.org/0000-0003-0392-3604
Danielle Lucero https://orcid.org/0000-0002-9288-9331
Bradley S. Frank https://orcid.org/0000-0003-3599-1521
Olivier Ilbert https://orcid.org/0000-0002-7303-4397
Luke J.M. Davies https://orcid.org/0000-0003-3085-0922
Ali A. Khostovan https://orcid.org/0000-0002-0101-336X
Mara Salvato https://orcid.org/0000-0001-7116-9303

## References

Barnes, D. G., Staveley-Smith, L., de Blok, W. J. G., et al. 2001, MNRAS, 322, 486
Behroozi, P., Wechsler, R. H., Hearin, A. P., & Conroy, C. 2019, MNRAS, 488, 3143
Bera, A., Kanekar, N., Chengalur, J. N., & Bagla, J. S. 2019, ApJL, 882, L7
Bera, A., Kanekar, N., Chengalur, J. N., & Bagla, J. S. 2023a, ApJ, 950, L18
Bera, A., Kanekar, N., Chengalur, J. N., & Bagla, J. S. 2023b, ApJL, 956, L15
Bigiel, F., Leroy, A., Walter, F., et al. 2010, AJ, 140, 1194
Blue Bird, J., Davis, J., Luber, N., et al. 2020, MNRAS, 492, 153
Bigiel, F., Leroy, A., Walter, F., et al. 2008, AJ, 136, 2846
Boselli, A., Cortese, L., Boquien, M., et al. 2014, A&A, 564, A67
Brown, T., Catinella, B., Cortese, L., et al. 2017, MNRAS, 466, 1275
Cantalupo, S., Arrigoni-Battaia, F., Prochaska, J. X., Hennawi, J. F., & Madau, P. 2014, Natur, 506, 63
Carilli, C. L., & Walter, F. 2013, ARAA, 51, 105
Cappellari, M., Emsellem, E., Krajnovic, D., et al. 2011, MNRAS, 413, 813
Catinella, B., Saintonge, A., Janowiecki, S., et al. 2018, MNRAS, 476, 875
Catinella, B., Schiminovich, D., Cortese, L., et al. 2013, MNRAS, 436, 34
Catinella, B., Schiminovich, D., Kauffmann, G., et al. 2010, MNRAS, 403, 683
Chabrier, G. 2003, PASP, 115, 763
Chen, Q., Meyer, M., Popping, A., et al. 2021, MNRAS, 508, 2
Chowdhury, A., Kanekar, N., & Chengalur, J. N. 2022, ApJL, 941, L6
Chowdhury, A., Kanekar, N., Chengalur, J. N., Sethi, S., & Dwarakanath, K. S. 2020, Natur, 586, 369
Chowdhury, A., Kanekar, N., Das, B., Dwarakanath, K. S., & Sethi, S. 2021, ApJL, 913, L24
Collier, J. D., Frank, B., Sekhar, S., & Taylor, A. R. 2021, in XXXIVth General Assembly and Scientific Symp. Int. Union of Radio Science (URSI GASS) (Piscataway, NJ: IEEE), 14
Comrie, A., Wang, K.-S., Hsu, S.-C., et al., 2021 CARTA: The Cube Analysis and Rendering Tool for Astronomy v2, Zenodo, doi:10.5281/zenodo.3377984
Conselice, C. J., Mortlock, A., Bluck, A. F. L., Grützbauch, R., & Duncan, K. 2013, MNRAS, 430, 1051
Cortese, L., Catinella, B., & Smith, R. 2021, PASA, 38, e035
Daddi, E., Rich, R. M., Valentino, F., et al. 2022, ApJL, 926, L21
Dahlen, T., Mobasher, B., Faber, S. M., et al. 2013, ApJ, 775, 93
Davies, L. J. M., Robotham, A. S. G., Driver, S. P., et al. 2018, MNRAS, 480, 768
de Blok, W. J. G., Healy, J., Maccagni, F. M., et al. 2024, A&A, 688, A109
Delhaize, J., Meyer, M. J., Staveley-Smith, L., & Boyle, B. J. 2013, MNRAS, 433, 1398
Dénes, H., Kilborn, V. A., Koribalski, B. S., & Wong, O. I. 2016, MNRAS, 455, 1294
DESI Collaboration, Adame, A. G., Aguilar, J., et al. 2024, AJ, 168, 58
Dodson, R., Momjian, E., Pisano, D. J., et al. 2022, AJ, 163, 59
Eibensteiner, C., Sun, J., Bigiel, F., et al. 2024, A&A, 691, A163
Elson, E. C., Baker, A. J., & Blyth, S. L. 2019, MNRAS, 486, 4894
Elson, E. C., Blyth, S. L., & Baker, A. J. 2016, MNRAS, 460, 4366
Evoli, C., Salucci, P., Lapi, A., & Danese, L. 2011, ApJ, 743, 45
Fabello, S., Catinella, B., Giovanelli, R., et al. 2011, MNRAS, 411, 993
Faisst, A. L., Capak, P., Hsieh, B. C., et al. 2016, ApJ, 821, 122
Fernández, X., Gim, H. B., van Gorkom, J. H., et al. 2016, ApJL, 824, L1
Fernández, X., van Gorkom, J. H., Hess, K. M., et al. 2013, ApJL, 770, L29
Forrest, B., Marsan, Z. C., Annunziatella, M., et al. 2020, ApJ, 903, 47
Geréb, K., Morganti, R., Oosterloo, T. A., Guglielmino, G., & Prandoni, I. 2013, A&A, 558, A54
Geréb, K., Morganti, R., Oosterloo, T. A., Hoppmann, L., & Staveley-Smith, L. 2015, A&A, 580, A43
Giovanelli, R., Haynes, M. P., Kent, B. R., et al. 2005, AJ, 130, 2598
Guo, H., Jones, M. G., Wang, J., & Lin, L. 2021, ApJ, 918, 53
Guo, H., Wang, J., Jones, M. G., & Behroozi, P. 2023, ApJ, 955, 57
Gupta, Y., Ajithkumar, B., Kale, H. S., et al. 2017, CSci, 113, 707
Haynes, M. P., Giovanelli, R., Kent, B. R., et al. 2018, ApJ, 861, 49
Healy, J., Blyth, S. L., Elson, E., et al. 2019, MNRAS, 487, 4901
Healy, J., Blyth, S. L., Verheijen, M. A. W., et al. 2021, A&A, 650, A76
Hess, K. M., Luber, N. M., Fernández, X., et al. 2019, MNRAS, 484, 2234
Hoppmann, L., Staveley-Smith, L., Freudling, W., et al. 2015, MNRAS, 452, 3726
Huang, S., Haynes, M. P., Giovanelli, R., & Brinchmann, J. 2012, ApJ, 756, 113
Ilbert, O., McCracken, H. J., Le Fèvre, O., et al. 2013, A&A, 556, A55
Jaffé, Y. L., Verheijen, M. A. W., Haines, C. P., et al. 2016, MNRAS, 461, 1202
Janowiecki, S., Catinella, B., Cortese, L., et al. 2017, MNRAS, 466, 4795
Jarvis, M., Taylor, R., Agudo, I., et al. 2016, in Proc. MeerKAT Science: On the Pathway to the SKA (Trieste: PoS)
Jarvis, M. J., Bonfield, D. G., Bruce, V. A., et al. 2013, MNRAS, 428, 1281
Jonas, J. & MeerKAT Team 2016, in Proc. MeerKAT Science: On the Pathway to the SKA (Trieste: PoS), 1
Karim, A., Schinnerer, E., Martínez-Sansigre, A., et al. 2011, ApJ, 730, 61
Kennicutt, R. C. J. 1998, ARA&A, 36, 189
Kleiner, D., Pimbblet, K. A., Jones, D. H., Koribalski, B. S., & Serra, P. 2017, MNRAS, 466, 4692
Lah, P., Pracy, M. B., Chengalur, J. N., et al. 2009, MNRAS, 399, 1447
Laigle, C., McCracken, H. J., Ilbert, O., et al. 2016, ApJS, 224, 24
Laudage, S., Eibensteiner, C., Bigiel, F., et al. 2024, A&A, 690, A169
Lilly, S. J., Le Fèvre, O., Renzini, A., et al. 2007, ApJS, 172, 70
Lin, L., Pan, H.-A., Ellison, S. L., et al. 2019, ApJL, 884, L33
Maccagni, F. M., Morganti, R., Oosterloo, T. A., Geréb, K., & Maddox, N. 2017, A&A, 604, A43
Madau, P., & Dickinson, M. 2014, ARA&A, 52, 415
Maddox, N., Frank, B. S., Ponomareva, A. A., et al. 2021, A&A, 646, A35
Maddox, N., Hess, K. M., Blyth, S. L., & Jarvis, M. J. 2013, MNRAS, 433, 2613
Maddox, N., Hess, K. M., Obreschkow, D., Jarvis, M. J., & Blyth, S. L. 2015, MNRAS, 447, 1610
Mancera Piña, P. E., Posti, L., Fraternali, F., Adams, E. A. K., & Oosterloo, T. 2021a, A&A, 647, A76
Mancera Piña, P. E., Posti, L., Pezzulli, G., et al. 2021b, A&A, 651, L15
McCracken, H. J., Milvang-Jensen, B., Dunlop, J., et al. 2012, A&A, 544, A156
Meyer, M. 2009, in Proc. Panoramic Radio Astronomy, wide-field 1-2 GHz research on galaxy evolution (Trieste: PoS)
Morselli, L., Renzini, A., Enia, A., & Rodighiero, G. 2021, MNRAS, 502, L85
Nan, R., Li, D., Jin, C., et al. 2011, IJMPD, 20, 989
Nelson, D., Springel, V., Pillepich, A., et al. 2019, ComAC, 6, 2
Obreschkow, D., & Meyer, M. 2014, arXiv:1406.0966
Oliver, S., Frost, M., Farrah, D., et al. 2010, MNRAS, 405, 2279
Padmanabhan, H., & Loeb, A. 2020, MNRAS, 496, 1124
Pan, H., Jarvis, M. J., Santos, M. G., et al. 2023, MNRAS, 525, 256
Peng, Y.-j., & Renzini, A. 2020, MNRAS, 491, L51
Popesso, P., Concas, A., Cresci, G., et al. 2023, MNRAS, 519, 1526
Rajohnson, S. H. A., Frank, B. S., Ponomareva, A. A., et al. 2022, MNRAS, 512, 2697
Rhee, J., Lah, P., Briggs, F. H., et al. 2018, MNRAS, 473, 1879
Rhee, J., Lah, P., Chengalur, J. N., Briggs, F. H., & Colless, M. 2016, MNRAS, 460, 2675
Rhee, J., Meyer, M., Popping, A., et al. 2023, MNRAS, 518, 4646
Rhee, J., Zwaan, M. A., Briggs, F. H., et al. 2013, MNRAS, 435, 2693
Roberts, M. S. 1962, AJ, 67, 437
Rodighiero, G., Daddi, E., Baronchelli, I., et al. 2011, ApJL, 739, L40
Rodighiero, G., Renzini, A., Daddi, E., et al. 2014, MNRAS, 443, 19
Romeo, A. B., Agertz, O., & Renaud, F. 2020, MNRAS, 499, 5656
Romeo, A. B., Agertz, O., & Renaud, F. 2023, MNRAS, 518, 1002

Roychowdhury, S., Chengalur, J. N., Begum, A., & Karachentsev, I. D. 2009, MNRAS, 397, 1435
Roychowdhury, S., Huang, M.-L., Kauffmann, G., Wang, J., & Chengalur, J. N. 2015, MNRAS, 449, 3700
Saintonge, A., & Catinella, B. 2022, ARA&A, 60, 319
Saintonge, A., Catinella, B., Tacconi, L. J., et al. 2017, ApJS, 233, 22
Sánchez Almeida, J., Elmegreen, B. G., Muñoz-Tuñón, C., & Elmegreen, D. M. 2014, A&ARv, 22, 71
Sancisi, R., Fraternali, F., Oosterloo, T., & van der Hulst, T. 2008, A&ARv, 15, 189
Schaye, J., Crain, R. A., Bower, R. G., et al. 2015, MNRAS, 446, 521
Schreiber, C., Glazebrook, K., Nanayakkara, T., et al. 2018, A&A, 618, A85
Scoville, N., Aussel, H., Brusa, M., et al. 2007, ApJS, 172, 1
Serra, P., Oosterloo, T., Morganti, R., et al. 2012, MNRAS, 422, 1835
Sinigaglia, F., Rodighiero, G., Elson, E., et al. 2022, ApJL, 935, L13
Sinigaglia, F., Rodighiero, G., Elson, E., et al. 2024, MNRAS, 529, 4192
Speagle, J. S., Steinhardt, C. L., Capak, P. L., & Silverman, J. D. 2014, ApJS, 214, 15
Tacconi, L. J., Genzel, R., & Sternberg, A. 2020, ARA&A, 58, 157
Tacconi, L. J., Genzel, R., Saintonge, A., et al. 2018, ApJ, 853, 179
Verheijen, M., van Gorkom, J. H., Szomoru, A., et al. 2007, ApJL, 668, L9
Walter, F., Carilli, C., Neeleman, M., et al. 2020, ApJ, 902, 111
Wang, J., Koribalski, B. S., Serra, P., et al. 2016, MNRAS, 460, 2143
Wang, T.-M., Magnelli, B., Schinnerer, E., et al. 2022, A&A, 660, A142
Weaver, J. R., Davidzon, I., Toft, S., et al. 2023, A&A, 677, A184
Weaver, J. R., Kauffmann, O. B., Ilbert, O., et al. 2022, ApJS, 258, 11
York, D. G., Adelman, J., Anderson, J. E. J., et al. 2000, AJ, 120, 1579
Zhou, Z., Wu, H., Zhou, X., & Ma, J. 2018, PASP, 130, 094101
Zwaan, M. A. 2000, PhD thesis, Rijksuniversiteit Groningen